%% file: EDosDraft.tex
\documentclass[onecolumn,11pt]{article}
\usepackage{jheppub}

\usepackage[font={small}]{caption}
\usepackage{subfigure}
\usepackage{graphicx}
\usepackage{dcolumn}

\usepackage{comment}

\usepackage{float}
\usepackage[normalem]{ulem}

\usepackage{mathrsfs} 

\usepackage{multicol}
\usepackage{cancel}
\usepackage{mathtools}
\usepackage{hyperref}

\usepackage{subfigure}
\usepackage{xcolor}
\usepackage{simplewick}

\usepackage{amsmath}
\usepackage{amsfonts}
\usepackage{amssymb}
\usepackage{graphicx}
\usepackage{simplewick}
\usepackage{hyperref}
\usepackage{multirow}
\usepackage{enumitem}
\usepackage{physics}
\usepackage{booktabs}
\usepackage{xcolor}
\usepackage{dsfont}
\usepackage{subfigure}

\def\({\left(} \def\){\right)}
\def\[{\left[} \def\]{\right]}

\newcommand{\eg}{{\it e.g.,}\ }
\newcommand{\ie}{{\it i.e.,}\ }

\def\Tr{{\rm Tr}}

\def\p{\partial}

\def\le{\left(}
\def\ri{\right)}

\newcommand{\cE}{\mathcal{E}}

\makeatletter
\DeclareRobustCommand{\cev}[1]{%
  \mathpalette\do@cev{#1}%
}
\newcommand{\do@cev}[2]{%
  \fix@cev{#1}{+}%
  \reflectbox{$\m@th#1\vec{\reflectbox{$\fix@cev{#1}{-}\m@th#1#2\fix@cev{#1}{+}$}}$}%
  \fix@cev{#1}{-}%
}
\newcommand{\fix@cev}[2]{%
  \ifx#1\displaystyle
    \mkern#23mu
  \else
    \ifx#1\textstyle
      \mkern#23mu
    \else
      \ifx#1\scriptstyle
        \mkern#22mu
      \else
        \mkern#22mu
      \fi
    \fi
  \fi
}

\makeatother

\newcommand{\be}{\begin{equation}}
	\newcommand{\ee}{\end{equation}}
\newcommand{\bea}{\begin{equation} \begin{aligned}}
	\newcommand{\eea}{\end{aligned} \end{equation}}

\def\le{\left(}
\def\ri{\right)}

\def\Re           {{\rm Re\hskip0.1em}}

\renewcommand{\eqref}[1]{(\ref{#1})}

\usepackage{changepage} 

\numberwithin{equation}{section}

\newcommand{\sri}   {\mathsf{S_{Ri}}}

\begin{document}

\title{Entanglement spectra from holography}

\author{Stefano Baiguera$^{1,2}$, Shira Chapman$^{3}$, Christian Northe$^{3}$, Giuseppe Policastro$^{4}$, Tal Schwartzman$^{3}$}

\affiliation{$^1$INFN Sezione di Perugia, Via A. Pascoli, 06123 Perugia, Italy}
\affiliation{$^2$Dipartimento di Matematica e Fisica, Universit\`{a} Cattolica del Sacro Cuore, \\ Via della Garzetta 48, 25133 Brescia, Italy}
\affiliation{$^3$ Department of Physics, Ben-Gurion University of the Negev, \\ David Ben Gurion Boulevard 1, Beer Sheva 84105, Israel \\
$^4$ Laboratoire de Physique  de l'\'Ecole Normale Sup\'eri{e}ure, \\
CNRS, PSL  Research University  and Sorbonne Universit\'es, \\
24 rue Lhomond, 75005 Paris, France}

\emailAdd{stefano.baiguera@pg.infn.it}
\emailAdd{schapman@bgu.ac.il}
\emailAdd{northe@post.bgu.ac.il}
\emailAdd{giuseppe.policastro@ens.fr}
\emailAdd{taljios@gmail.com}

\abstract{ \sloppy
The entanglement spectrum of a bipartite quantum system is given by the distribution of eigenvalues of the modular Hamiltonian.
In this work, we compute the entanglement spectrum in the vacuum state for a subregion of a $d$-dimensional conformal field theory (CFT) admitting a holographic dual.
In the case of a spherical (or planar) entangling surface, we recover known results in two dimensions, including the Cardy formula in the high energy regime.
In higher dimensions $d>2$, we analytically determine a generalization of the Cardy formula valid at large energies and consistent with previous studies of 
CFT spectra in the literature. 
We also investigate numerically the spectrum at energy levels far above the modular ground state energy.
We extend our analysis to the supersymmetric point of Einstein-Maxwell gravity, providing exact results when $d=2,3$, and a generalization of the Cardy formula at high energies in generic dimension $d$. 
We consider small shape deformations of a spherical entangling surface, for both the non-supersymmetric and the supersymmetric cases. In all  cases we find that the high-energy scaling of the microcanonical entropy with the modular energy is unaffected by the shape deformation.
This result suggests that the high-energy regime of the entanglement spectra carries universal information, independent of the shape of the entangling surface.}

\maketitle

\input{Draft_Sections/Intro}

\input{Draft_Sections/HolographicRenyi}

\input{Draft_Sections/HolographicEDOS}

\input{Draft_Sections/Susy_DOS}

\input{Draft_Sections/ShapeDeformations}

\input{Draft_Sections/Conclusion}


\section*{Acknowledgements}

We are happy to thank Roberto Auzzi, Lorenzo Bianchi, Dami\'{a}n A.~Galante, Kausik Ghosh, Giuseppe Nardelli, Erik Tonni and Itamar Yaakov for valuable discussions.
We especially thank Alexandre Belin and Marco Meineri for insightful discussions and comments on the draft.
The authors gratefully acknowledge the organizers and participants of the workshop “Defects, from condensed matter to quantum gravity” in Pollica (Italy) for hospitality and for interesting discussions.
The work of SB was supported by the INFN grant ``Gauge Theories and Strings'' (GAST) via a research grant on ``Holographic dualities, quantum information and gravity''.
The work of SB, SC, CN and TS was supported by the Israel Science Foundation (grant No. 1417/21), by the German Research Foundation through a German-Israeli Project Cooperation (DIP) grant “Holography and the Swampland”, by Carole and Marcus Weinstein through the BGU Presidential Faculty Recruitment Fund, by the ISF Center of Excellence for theoretical high energy physics.
SC and TS are also supported by the ERC starting Grant dSHologQI (project number 101117338).

\appendix

\input{Draft_Sections/App_Laplace}
\input{Draft_Sections/App_Universality}


\addcontentsline{toc}{section}{References}

\bibliography{biblio}
\bibliographystyle{biblio}

\end{document}

%% file: Draft_Sections/Intro.tex
\section{Introduction}

Quantum entanglement is a fundamental phenomenon in quantum mechanics that has intrigued and puzzled scientists since it was first conceptualized. The study of entanglement has profound implications for our understanding of information processing, quantum matter and even gravity. Most strikingly, it challenges the classical notion of locality, allowing for conceptual leaps such as ER=EPR (Einstein-Rosen=Einstein-Podolsky-Rosen) \cite{Maldacena:2013xja}.

Traditionally, entanglement is often quantified by entanglement entropy. However, much more information is contained in the so-called entanglement spectrum \cite{PhysRevLett.101.010504}. 
It consists of the eigenvalues of the reduced density matrix, and serves as a fingerprint of the entanglement properties within two complementary subsystems. Remarkably, the entanglement spectrum can reveal novel phase transitions, topological order, and critical phenomena that are not always apparent through traditional means \cite{PhysRevLett.103.016801,PhysRevLett.101.010504,PhysRevLett.104.130502,PhysRevLett.104.156404,PhysRevLett.105.080501,PhysRevLett.107.157001,PhysRevLett.105.077202,PhysRevLett.108.196402,PhysRevB.84.205136,Metlitski:2011pr,PhysRevLett.108.227201,PhysRevLett.116.190401,PhysRevB.83.045110,PhysRevLett.115.267206,PhysRevB.93.174202,PhysRevLett.117.160601}. So far, much of our understanding of  the entanglement spectrum has been gathered in low-dimensional systems \cite{Cardy:1986ie,Strominger:1997eq,Hartman:2013mia,Hartman:2014oaa}. One aim of our present work is to cover ground in higher dimensions.

The non-local nature of entanglement allows to naturally interpret it as "glue" of space itself, an idea concretely realized within the AdS/CFT correspondence by the Ryu-Takayanagi formula \cite{Ryu:2006bv,Ryu:2006ef}. 
While many properties of entanglement entropy and its cousins have been investigated \cite{Hubeny:2007xt,Nishioka:2009un,Casini:2011kv,Hung:2011nu,Hung:2011xb,Myers:2012ed,Belin:2013dva,Belin:2013uta,Dong:2013qoa,Galante:2013wta,Faulkner:2013ana,Rangamani:2016dms,Dong:2016fnf}, the entanglement spectrum of holographic theories has not been addressed so far. 
It is the main focus of this paper to fill this gap; in particular, our analysis is carried out in manifolds with arbitrary spacetime dimension.

\vskip 2mm
\noindent
In the case of a conformal field theory, the problem of finding the entanglement spectrum can, in some cases, be related to the better-studied problem of finding the energy spectrum of the CFT (with boundaries). 
The benchmark for a universal result for CFT spectra is provided by the Cardy formula, which describes the asymptotic density of states $D(E)$ in any unitary two-dimensional CFT as \cite{Cardy:1986ie,carlip1998we,Mukhametzhanov:2019pzy} 
\be
S(E) = \log D(E) \approx 2\pi \sqrt{\frac{c}{6} \le E - \frac{c}{24} \ri } \, , 
\label{eq:Cardy_intro}
\ee
where $S(E)$ denotes the micro-canonical entropy of states at fixed energy $E$ and $c$ is the central charge in two dimensions.
The regime of validity of this formula is
\be
\text{$c$ fixed}\, , \qquad
E \gg E_0 \, ,
\label{eq:validity1_Cardy_intro}
\ee
where $E_0=c/24$ is the Casimir (or vacuum) energy.
We can get remarkable insights on this result from the study of CFTs with holographic dual. Here, the central charge is mapped to the AdS radius and Newton's constant via Brown-Henneaux's relation, and the entropy entering Cardy's formula \eqref{eq:Cardy_intro} is related to the horizon area of a BTZ black hole with two asymptotic and disconnected boundaries \cite{Strominger:1997eq}.
One advantage of holography is that it allows us to explore the regime of validity differing from \eqref{eq:validity1_Cardy_intro}, namely
\be
c \gg 1 \, , \qquad 
E \gtrsim c \, ,
\label{eq:validity2_Cardy_intro}
\ee
The regime \eqref{eq:validity2_Cardy_intro} is interesting due to its connection to Bekenstein-Hawking entropy.
Here the density of states is of Cardy-type \eqref{eq:Cardy_intro} for energies $E>c/12$ \cite{Hartman:2014oaa}. 

The universality of Cardy's formula has been shown to extend to entanglement spectra in  \cite{Calabrese:2008iby, Alba:2017bgn}.  The same equation ~\eqref{eq:Cardy_intro}, albeit with $E$ interpreted as modular energy, was found to hold in the regime \eqref{eq:validity1_Cardy_intro}, while the holographic regime \eqref{eq:validity2_Cardy_intro} remains unexplored, a gap that is addressed in this paper. 

The existence of a higher-dimensional universal analog of the Cardy formula is less understood.
Such a generalization was first advanced in the context of CFT spectra by Verlinde, based on the holographic principle applied to a radiation-dominated closed FRW universe \cite{Verlinde:2000wg}.
Later on, this idea was refined and extended to other higher-dimensional cases either by using modular forms \cite{Shaghoulian:2015kta}, or holography in AdS Schwarzschild/Kerr black hole backgrounds \cite{Shaghoulian:2015lcn}, or focusing on free CFTs \cite{Behan:2012tc}.
In parallel, novel versions of the Cardy formula in higher dimensions were obtained from the high-temperature approximation of the partition function in supersymmetric theories \cite{DiPietro:2014bca,ArabiArdehali:2015iow,Zhou:2015cpa}.
Finally, the thermal effective action technique, often used in the hydrodynamics literature, was employed to extract the density of states in \cite{Bhattacharyya:2007vs} and in the recent work \cite{Benjamin:2023qsc}, leading to the following scaling behaviour:
\be
S(E) \sim \le  E-E_0 \ri^{\frac{d-1}{d}} \, .
\label{eq:Ooguri_intro}
\ee
When the spin of the dual operators is negligible, meaning that the spin is much smaller than the energy, $J \ll E$, the previous formula holds in the regime
\be
\text{$C_T$ fixed}\, , \qquad
E \gg E_0 \, ,
\label{eq:validity1_higherd_Cardy}
\ee
where $C_T$ is the higher-dimensional central charge controlling the two-point function of the energy-momentum tensor \cite{Osborn:1993cr,Erdmenger:1996yc} and $E_0$ is the lowest energy in the spectrum.
In other words, this window corresponds to the higher-dimensional generalization of the condition \eqref{eq:validity1_Cardy_intro}.
One of the purposes of the present paper is to explore the energetic regime
\be
C_T \gg 1 \, , \qquad
E \gtrsim C_T \, .
\label{eq:validity2_higherd_Cardy}
\ee
in the context of entanglement spectra by use of holography. This regime provides the generalization of eq.~\eqref{eq:validity2_Cardy_intro} to dimensions larger than two.

\paragraph{Summary of results.}
While the study of the regime \eqref{eq:validity2_higherd_Cardy} was initiated in \cite{Benjamin:2023qsc} for the case of standard CFT spectra, here we extend this line of research to entanglement spectra in CFTs with large central charge. Hence, $E$ is interpreted here as modular energy. Our main input is given by the holographic R\'{e}nyi entropies, that are related to the partition function via the replica trick \cite{Casini:2011kv,Hung:2011nu,Belin:2013uta}. 
In general number of dimensions, we show that the density of states takes the form
\be 
\label{eq:DE_delta_conclusions}
D(E) = \delta(E-E_0) + \Theta (E-E_0) \mathfrak{D}(E) \, ,
\ee
where $E_0$ is the lowest energy in the spectrum of the modular Hamiltonian.
In other words, there is always a divergent contribution around $E=E_0$ controlled by a Dirac $\delta$-distribution, while the relevant and physical information comes from energy states with $E>E_0$.
The computation of the latter contribution $\mathfrak{D}(E)$, simply denoted with $D(E)$ below, is the main goal of this work.

Quantitatively, we find the same result as \eqref{eq:Ooguri_intro} for the vacuum state in a (non-supersymmetric) CFT with a spherical entangling surface. This may be expected at first sight, since a conformal map, as discussed below, maps the entanglement problem into a thermal CFT state, albeit on hyperbolic space. 
The scaling of the density of states with the energy in general dimensions was also proposed as an ansatz in \cite{Hung:2011nu}.
From the outset, it is however not clear how the additional UV divergences riddling entanglement studies enter in this picture. These divergences need to be carefully accounted for in order to extract the universal structure of the density of states of entanglement spectra. Our results not only show how to incorporate said divergences  -- thereby deriving the scaling \eqref{eq:Ooguri_intro} from first principles -- but we also develop a new approach, which is easily exploited in two novel applications in this work, namely in supersymmetric theories and states with deformed entangling surfaces. 
Let us point out that further differences between regular CFT spectra and entanglement spectra arise by treating the entangling boundaries in the BCFT approach -- we leave this for future work however.

Let us turn to the first such novelty of our work, \ie the investigation of the entanglement density of states associated with the reduced density matrix of a supersymmetric CFT admitting a holographic dual.
In this case, one needs to select a $\mathrm{U}(1)$ subgroup of the global R-symmetry, and switch on a conjugate background gauge field that generates a non-vanishing Aharanov-Bohm flux across loops encircling the entangling surface. By appropriately tuning this flux, one can preserve supersymmetry and define a corresponding R\'{e}nyi entropy \cite{Nishioka:2013haa,Nishioka:2014mwa,Belin:2013uta}.
In this case, we show that the microcanonical entropy at high energies presents a universal scaling 
\be
S(E) \sim \begin{cases}
    \sqrt{E-E_0} & \text{if $d=2,3$} \\
    (E-E_0)^{\frac{d-2}{d-1}} & \text{for $d>3$}
\end{cases}
\label{eq:SUSY_result_intro}
\ee
which differs from eq.~\eqref{eq:Ooguri_intro} when $d>2$.
Moreover, supersymmetry allows us to compute the exact density of states in dimensions $d=2,3$, as summarized by the identity 
\be
D(E)\Big|_{\rm SUSY}^{d=2,3} = \sqrt{\frac{\kappa}{E-E_0}} e^{\eta} \, I_1 \le 2 \sqrt{\kappa(E-E_0)} \ri \, , 
\label{eq:SUSY_result_intro2}
\ee
where $\kappa, \eta$ are certain constants (see section \ref{ssec:exact_SUSYDOS_2d3d} for details), and $I_1$ is the modified Bessel function of the first kind.
While the two-dimensional case was already known \cite{Calabrese:2008iby}, our result for three dimensions is novel. Surprisingly, at high energies it shows the same scaling as the Cardy formula \eqref{eq:Cardy_intro}.

The results \eqref{eq:Ooguri_intro} and \eqref{eq:SUSY_result_intro} use as an input the R\'{e}nyi entropies computed for a spherical (or planar) entangling surface.
As long as a small deformation of this symmetric shape is performed, it is still possible to employ holography to compute the leading variation of the R\'{e}nyi entropy \cite{Dong:2016wcf,Bianchi:2016xvf,Baiguera:2022sao}. 
In this work, we will show that a shape deformation only affects certain details in the prefactors of the density of states, but it does not modify the scaling with energy.
This provides evidence for a novel kind of universality in the generalized Cardy formula, since shape deformations have no counterpart in standard CFT spectra.

\paragraph{Outline.}
The paper is organized as follows.
We review in section~\ref{ssec:review_holo_Renyi} the main results for the R\'{e}nyi entropy of a holographic CFT, focusing in particular on the cases where the dual hyperbolic black hole is uncharged or enjoys a supersymmetric enhancement.
After considering a setting with spherical (or planar) entangling surface, we also discuss the case where a small deformation of its shape is performed.
Section~\ref{sec:dual_DOS} is the core of this paper, since it presents the general strategy to compute the density of states using the R\'{e}nyi entropy as an input.
We then focus in section~\ref{sec:holographic_CFT_uncharged} on the uncharged case and find a universal high-energy scaling of the microcanonical entropy valid in any dimension, thus generalizing the Cardy formula. Our results are supported by a numerical analysis and explicit examples.
We consider the case of supersymmetric Einstein-Maxwell gravity and its dual CFT in section~\ref{subsec:DOS_SUSY}, where we perform a similar analysis.
In this case, we also find exact results (valid at any energy scale) in two and three dimensions.
The density of states in the presence of a shape deformation of the entangling surface is reserved to section~\ref{sec:shape_def_DOS}.
Finally, we discuss the implications and several possible extensions of our results in section~\ref{sec:conclusions}.
Appendix~\ref{app:inverse_Laplace} contains technical details on the inverse Laplace transform and the saddle point approximation.
Appendix~\ref{app:universality_Edep} argues that the results obtained in this work are universal, \ie independent of the regularization scheme.

%% file: Draft_Sections/HolographicRenyi.tex
\section{Review of the holographic R\'{e}nyi entropy}
\label{ssec:review_holo_Renyi}

In this section, we review the main steps characterizing the holographic computation of R\'{e}nyi entropy, and summarize the results relevant for the later sections.
We begin with the general definitions and the introduction of the replica trick in section~\ref{ssec:replica_trick}.
The problem of computing R\'{e}nyi entropies in a CFT with a spherical (or planar) entangling surface is then mapped to a thermal setting in hyperbolic space in section~\ref{ssec:map_hyperbolic_space}. 
The relation to holography and the explicit expressions of the R\'{e}nyi entropies are discussed in section~\ref{ssec:holo_Renyi}.
Finally, in section~\ref{ssec:shape_def_review} we examine the case of performing a small shape deformation to the entangling surface.
For practical convenience, we present the general framework where the physical system admits a global charge \cite{Belin:2013uta,Baiguera:2022sao}.
Subsequently, we reduce to the special cases where either the conserved charge vanishes \cite{Casini:2011kv,Hung:2011nu,Mezei:2014zla,Allais:2014ata,Bianchi:2015liz,Dong:2016wcf,Bianchi:2016xvf}, or where supersymmetry (SUSY) is imposed \cite{Nishioka:2013haa,Nishioka:2014mwa}.

\subsection{Replica trick}
\label{ssec:replica_trick}

Consider a pure state described by a density matrix $\rho$ on a fixed time slice of a $d$--dimensional CFT (without loss of generality, we take the surface to be $t=0$).
A codimension-two entangling surface $\Sigma$ divides this spacelike slice into a subregion $A$ and its complement $\bar{A}$, leading to an associated factorization of the Hilbert space:
\begin{equation}
    \mathcal{H}\to\mathcal{H}_A\otimes \mathcal{H}_{\bar{A}} \, .
\end{equation}
This splitting defines a reduced density matrix $\rho_A = \Tr_{\bar{A}} \rho$.
We assume that the CFT is invariant under a global $\mathrm{U}(1)$ symmetry associated with a conserved charge $Q$ that commutes with the density matrix, \ie $[\rho, Q]=0$. Consequently, $[\rho_A, Q_A]=0$, where $Q_A$ is the charge operator restricted to the subregion $A$.

The replica trick is a technique used to compute R\'{e}nyi entropies in terms of the (Euclidean) path integral over a replicated geometry, obtained by taking  $n$ copies of the field theory, glued together with appropriate boundary conditions \cite{Calabrese:2004eu,Calabrese:2005zw}.
The key identity is:
\be
\Tr \left[ \rho_A \, \frac{e^{\mu Q_A}}{n_A (\mu)} \right]^n = 
\frac{Z_n (\mu)}{\le Z_1 (\mu) \ri^n} \,,
\label{eq:general_replica_trick}
\ee
where $Z_n(\mu)$ denotes the grand-canonical partition function evaluated on an $n$--fold cover of flat space, with branch cuts along the subregion $A$.
In this expression, $\mu$ is a chemical potential conjugate to the charge $Q_A$. 
In terms of a background gauge potential $B_{\mu}$ coupled to the relevant conserved current $j^{\mu}$, the chemical potential corresponds to the Aharonov-Bohm flux across any loop $\mathcal{C}$ encircling the entangling surface via $\oint_{\mathcal{C}} B = -i n \mu$.
Finally, the constant $n_A(\mu) = \Tr \le \rho_A e^{\mu Q_A} \ri$ is chosen such that the expression \eqref{eq:general_replica_trick} is normalized to 1 as $n \rightarrow 1 $, for any value of $\mu$.

In terms of the partition function, the charged R\'{e}nyi entropies are defined by \cite{Belin:2013uta}
\begin{equation}
S_n(\mu) := \frac{1}{1-n} \log \Tr \left[\rho_A\frac{e^{\mu Q_A}}{n_A(\mu)}\right]^n 
= \frac{1}{1-n}\left(\log Z_n(\mu) - n \log Z_1(\mu) \right) \, ,
\label{eq:def_charged_Renyi}
\end{equation}
where the identity \eqref{eq:general_replica_trick} was used in the last step.
This expression reduces to the usual R\'{e}nyi entropy when $\mu=0$ \cite{Casini:2011kv,Hung:2011nu}, while it becomes a supersymmetric R\'{e}nyi entropy when the chemical potential takes the special form $\mu\propto \frac{n-1}{n}$ for the classes of theories we analyze below \cite{Nishioka:2013haa,Nishioka:2014mwa}.
In the latter case, the conserved charge $Q$ needs to belong to a $\mathrm{U}(1)$ subgroup of the R-symmetry.

\subsection{Map to hyperbolic space}
\label{ssec:map_hyperbolic_space}

Let us assume that the CFT is in the vacuum state of flat space, and that the entangling surface $\Sigma$ has a spherical shape with radius $R$.\footnote{A similar argument applies if the entangling surface is planar, since this shape is related to the sphere via a conformal transformation.}
In order to simplify the evaluation of entanglement in the subregion $A$, it is useful to map the system to a thermal state on the hyperbolic cylinder $\mathbb{R} \times H^{d-1}$ \cite{Casini:2011kv}.
One begins with Minkowski space in polar coordinates
\be
ds^2 = -dt^2 + dr^2 + r^2 d\Omega_{d-2}^2 \, ,
\label{eq:polar_Minkowski}
\ee
where $d\Omega_{d-2}^2$ is the line element of the unit sphere $S^{d-2}$.
By performing the following change of coordinates\footnote{Notice that in the case of a planar entangling surface, a length scale $R$ still needs to be included in the system to make the arguments of the hyperbolic functions dimensionless. The difference in the spherical case is that $R$ has a direct interpretation as the radius of the sphere. In principle one could have two different scales, one associated to $\tau$ and another to $t,r$. Here, they are chosen to be equal for convenience.}
\be
t = R \, \frac{\sinh \le \frac{\tau}{R} \ri}{\cosh u + \cosh \le \frac{\tau}{R} \ri} \, , \qquad
r = R \, \frac{\sinh u}{\cosh u + \cosh \le \frac{\tau}{R} \ri} \, , 
\label{eq:CHM_coord_transf}
\ee
the metric \eqref{eq:polar_Minkowski} becomes
\be
ds^2 = \Omega^2 \left[ - d\tau^2 + R^2 \le du^2 + \sinh^2 u \, d\Omega_{d-2}^2 \ri   \right] \, , \qquad
\Omega := \frac{1}{\cosh u + \cosh \le \frac{\tau}{R} \ri} \, .
\label{eq:CHM_hyperbolic_space}
\ee
After a Weyl transformation which removes the prefactor $\Omega^2$, the geometry in eq.~\eqref{eq:CHM_hyperbolic_space} is recognized to be $\mathbb{R} \times H^{d-1}$, where $R$ is the curvature radius of the hyperbolic space.
Notice that in this coordinate system, the original entangling surface has been pushed all the way to the asymptotic boundary, since $(t,r)=(0,R)$ corresponds to $u \rightarrow \infty$.
Furthermore, one can show that the coordinate transformation \eqref{eq:CHM_coord_transf} maps the causal development $\mathcal{D}(A)$ of the subregion $A$ (delimited by the spherical entangling surface $\Sigma$) to the full hyperbolic space.

\begin{figure}[ht]
    \centering
    \includegraphics[scale=0.3]{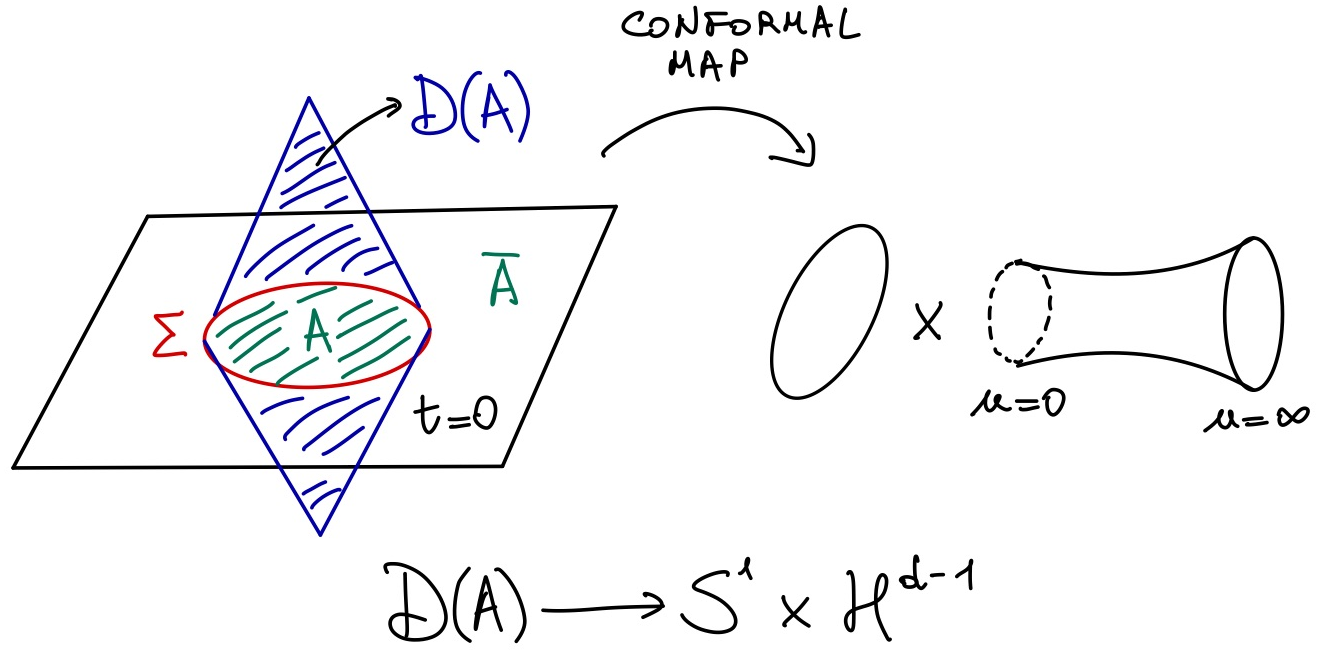}
    \caption{Map between the causal development $\mathcal{D}(A)$ inside a spherical entangling surface in Minkowski space to hyperbolic space $S^1 \times H^{d-1}$.}
    \label{fig:CHM}
\end{figure}

In Euclidean signature, the requirement to avoid conical singularities after the coordinate transformation \eqref{eq:CHM_coord_transf} implies the following periodicity of the time coordinate: 
\be
\tau_E \sim \tau_E + 2 \pi R \, .
\label{eq:periodicity}
\ee
This finally leads to the geometry $S^1 \times H^{d-1}$, as depicted in fig.~\ref{fig:CHM}.
The periodicity of the Euclidean time generated an effective temperature $T_0 = (2\pi R)^{-1}$ in the system. 

The crucial point of the above mapping is that the reduced density matrix $\rho_A$ on the causal development $\mathcal{D}(A)$ is related to the thermal density matrix $\rho_{\rm therm}$ on the hyperbolic space as follows \cite{Belin:2013uta}
\begin{subequations}
\be
\rho_A \, \frac{e^{\mu Q_A}}{n_A (\mu)} = U^{-1} \rho_{\rm therm} U \, , 
\label{eq:map_rhoA_rhoterm}
\ee
\be
\rho_A = e^{-K} \, ,
\qquad \rho_{\rm therm} = \frac{e^{-H/T_0 + \mu Q}}{Z(T_0, \mu)} \, ,
\ee
\end{subequations}
where $K$ is the modular Hamiltonian of the CFT, $H$ the Hamiltonian of the thermal system in the hyperbolic space, $Z(T_0, \mu)$ the grand-canonical partition function.
In eq.~\eqref{eq:map_rhoA_rhoterm}, $U$ is the unitary operator implementing the conformal transformation between the two geometries.

After performing the replica trick, the Euclidean time coordinate acquires periodicity
\be
\tau_E \sim \tau_E + 2 \pi R n \, ,
\label{eq:replica_periodicity_tau}
\ee
because the geometry is an $n$-fold cover of flat space. Consequently, the effective temperature becomes $T_0/n=(2\pi R n)^{-1}$.
Due to cyclicity property of the trace, the R\'{e}nyi entropies \eqref{eq:def_charged_Renyi} are insensitive to unitary transformations.
Therefore, using the map \eqref{eq:map_rhoA_rhoterm}, the R\'{e}nyi entropies can be computed in terms of the thermal grand-partition function $Z$ on the hyperbolic space as
\be
S_n (\mu) = \frac{1}{1-n} \left(\log Z (T_0/n, \mu) - n \log Z (T_0, \mu) \right) \, .
\label{eq:Renyi_as_therm_partition}
\ee

\subsection{Holographic R\'{e}nyi entropies}
\label{ssec:holo_Renyi}

It is usually difficult to compute the R\'{e}nyi entropies \eqref{eq:Renyi_as_therm_partition} for an arbitrary $d$--dimensional CFT.
However, in a holographic CFT further progress is made by exploiting the fact that the partition function is related to the regularized on-shell gravitational action $I_0$ via $I_0 = - \log Z(T_0,\mu)$.
The key point is to find a black hole solution in asymptotically AdS space whose asymptotic boundary is given by the hyperbolic geometry $S^1 \times H^{d-1}$ obtained in section~\ref{ssec:map_hyperbolic_space}.
In the context of the charged R\'{e}nyi entropies \eqref{eq:def_charged_Renyi}, the existence of a global charge on the CFT side
implies that the bulk theory includes a gauge field dual to the conserved current, and that the black hole is charged.

We restrict to the case of Einstein-Maxwell gravity in $d+1$ dimensions.
The Euclidean action reads \cite{Chamblin:1999tk}
\be
I_{\rm EM} = - \frac{1}{2 \ell_P^{d-1}} 
\int d^{d+1} x \, \sqrt{g} \le \mathcal{R} + \frac{d(d-1)}{L^2} - \frac{\ell_*^2}{4} F_{\mu\nu} F^{\mu\nu}  \ri \, ,
\label{eq:EM_action}
\ee
where $\ell_P$ is the Planck length, $L$ the AdS curvature radius, and $\ell_*$ is a coupling constant for the gauge field.\footnote{\label{foot:Planck_length} The Planck length is related to Newton's gravitational constant via $8\pi G_N = \ell_P^{d-1}$.}
The equations of motion admit as solutions charged topological black holes with metric \cite{Birmingham:1998nr,Cai:1998vy,Cai:2001jc,Cai:2004pz}
\be
ds^2 =  G(r) \frac{L^2}{R^2} d \tau^2 + \frac{dr^2}{G(r)} + r^2 d\Sigma^2_{d-1} \, ,
\label{eq:metric_BH}
\ee
where $\tau$ is the Euclidean time coordinate, $R$ is the curvature scale of the hyperbolic slices, and $d \Sigma^2_{d+1} = du^2 + \sinh^2 u \, d\Omega^2_{d-1} $ is the metric on the hyperbolic space $H^{d-1}$ with unit curvature radius.
The blackening factor reads
\be
G(r) = \frac{r^2}{L^2} -1 - \frac{m}{r^{d-2}} + \frac{Q^2}{r^{2(d-2)}} \, ,
\label{eq:blackening_factor_Einstein_Maxwell}
\ee
where $Q$ is the black hole charge and $m$ an integration constant, which we call \textit{mass parameter}.
It characterizes the energy density of the black hole and is related to the black hole's asymptotic mass $M$ as follows: 
\be
M = \omega_d \, m \, , \qquad 
\omega_d := \frac{(d-1) V_{\Sigma}}{2 \ell_P^{d-1}} \, ,
\ee
where $V_{\Sigma}$ is the regularized dimensionless volume of the hyperbolic sections $H^{d-1}$. 
By performing a change of variables $y = \sinh u$ and introducing a UV cutoff $\delta$ such that the maximum radius in the hyperbolic geometry reads $y_{\rm max} = R/\delta$, the volume $V_{\Sigma}$ is written as \cite{Hung:2011nu}
\be
V_{\Sigma} =  \Omega_{d-2} \int_{1}^{y_{\rm max}} dy \, (y^2-1)^{\frac{d-3}{2}} 
= \frac{\Omega_{d-2}}{d-2} \left[  \le \frac{R}{\delta} \ri^{d-2}
-  \frac{(d-2)(d-3)}{2(d-4)} \le \frac{R}{\delta} \ri^{d-4}
+ \dots \right]  \, ,
\label{eq:def_VSigma}
\ee
where $\Omega_{d-2}$ denotes the area of the unit $(d-2)$-sphere.
The dots contain additional singular terms in the series expansion around $\delta=0$ with powers decreasing in steps of 2, until a logarithmic divergence (in even dimensions $d$) or a finite part (in odd dimensions) is reached.
Additional terms in the series vanish when the limit $\delta \rightarrow 0$ is taken.

Hyperbolic black holes in asymptotically AdS spacetime can have a negative mass, with lowest value $m_{\rm cr}$ determined by \cite{Cai:2004pz}
\be
m_{\rm cr} = -2 r_{\rm cr}^{d-2} \left[ 1 - \frac{(d-1) r_{\rm cr}^2}{(d-2) L^2}  \right] <0 \, , \qquad
r_{\rm cr}^2 = L^2 \, \frac{d-2}{d} \le 1 + \frac{Q^2}{r_{\rm cr}^{2(d-2)}} \ri \, ,
\label{eq:critical_mass_radius}
\ee
where $r_{\rm cr}$ is a critical radius.
When the mass parameter satisfies $m > m_{\rm cr}$, the blackening factor \eqref{eq:blackening_factor_Einstein_Maxwell} admits two distinct roots $r_{\pm}$, whose largest one (denoted by $r_h := r_+$) satisfies the following identity between the mass parameter and the black hole charge:
\be
m = \frac{r_h^{d-2}}{L^2} \le r_h^2 - L^2 \ri + \frac{Q^2}{r_h^{d-2}}
\label{eq:relation_mass_charge} \, .
\ee
The two event horizons $r_{\pm}$ coincide and the black hole becomes extremal (with vanishing temperature) when $m=m_{\rm cr}$.
In the regime $m < m_{\rm cr}$, the black hole has a naked singularity, without any event horizon.
In this work, we restrict to the range $m \geq m_{\rm cr}$.

The charged black hole \eqref{eq:metric_BH} is equipped with the following $\mathrm{U}(1)$ gauge field 
\be
A = -i \le  \sqrt{\frac{2(d-1)}{d-2}}  \frac{L Q}{R \ell_* \, r^{d-2}} -  \frac{\mu}{2 \pi R} \ri d\tau \, ,
\label{eq:zeroth_order_solution_gauge}
\ee
where $\mu$ is a chemical potential fixed by requiring that the gauge field vanishes at the horizon:
\be
\mu = 2 \pi \sqrt{\frac{2(d-1)}{d-2}}  \frac{L Q}{\ell_* \, r_h^{d-2}}  \, .
\label{eq:chemical_potential_and_charge}
\ee
In Einstein-Maxwell gravity, the thermal entropy is given by the Bekenstein-Hawking formula
\be
S_{\rm therm}  = \frac{2\pi}{\ell_P^{d-1}} V_{\Sigma} \, r_h^{d-1} \, .
\label{eq:thermal_entropy_BH}
\ee
The Hawking temperature reads
\be
T(r_h, \mu) = \frac{T_0}{2} L \, G'(r_h) = 
\frac{T_0  L}{2 r_h}  \left[ d \, \frac{r_h^2}{L^2} - (d-2) - \frac{(d-2)^2}{2(d-1)} \le \frac{\mu \ell_*}{2 \pi L} \ri^2  \right]  \, ,\label{temperature_EM}
\ee
where $T_0 = (2\pi R)^{-1}$ is the temperature of AdS Rindler space, corresponding to the periodicity \eqref{eq:periodicity} of the time coordinate.

In order to study R\'{e}nyi entropies, we apply the replica trick described in section~\ref{ssec:replica_trick} on the CFT side.
This changes the periodicity of the time coordinate to \eqref{eq:replica_periodicity_tau}, and the temperature to $T(r_h, \mu)= T_0/n$.
These requirements impose a relation between the dimensionless horizon radius $x_n = r_h/L$ and the replica index $n$, which reads
\be
x_n = \frac{1}{d \, n} + \sqrt{\frac{1}{d^2 n^2} + \frac{d-2}{d} +  \frac{(d-2)^2}{2d(d-1)} \le \frac{\mu \ell_*}{2 \pi L} \ri^2 } \, .
\label{eq:definition_xn_charged}
\ee
Consequently, all the physical quantities associated with the black hole acquire a dependence on $n$, and it is possible to extract the R\'{e}nyi entropies by using the identity
\begin{equation}
S_n(\mu)=\frac{n}{1-n} \frac{1}{T_0} \left[ \mathcal{G}(T_0) -\mathcal{G} \le \frac{T_0}{n} \ri \right] \, ,
\label{eq:Renyi_from_grand_potential}
\end{equation}
obtained from eq.~\eqref{eq:Renyi_as_therm_partition} by introducing the grand-potential $\mathcal{G}= -T_0 \log Z(T_0, \mu)$.
In this way, one finally obtains \cite{Belin:2013uta}
\begin{equation}
    S_n(\mu) = \pi V_\Sigma \le \frac{L}{\ell_{\mathrm{P}}} \ri^{d-1}
    \frac{n}{n-1}
\left[  \le  1+ \frac{d-2}{2(d-1)} \le \frac{\mu \ell_*}{2 \pi L} \ri^2 \ri  \le x_1^{d-2} -x_n^{d-2}  \ri + x_1^d - x_n^d 
\right] \, .
\label{eq:charged_Renyi_holo}
\end{equation}
First, one observes that the dependence on the chemical potential $\mu$ in eqs.~\eqref{eq:definition_xn_charged} and \eqref{eq:charged_Renyi_holo} drops out for $d=2$.
Moreover, it should be noticed that the charged R\'{e}nyi entropies \eqref{eq:charged_Renyi_holo} are always proportional to the volume $V_{\Sigma}$ of the hyperbolic space, which was shown to be divergent in eq.~\eqref{eq:def_VSigma}.
However, it is well-known that the power-law divergences
are scheme-dependent, as can be seen by rescaling the UV cutoff \cite{Rangamani:2016dms,Ryu:2006ef,Nishioka:2009un}.
The universal contribution to $V_{\Sigma}$ corresponds to either the logarithmic (in even dimensions) or the finite part (in odd dimensions), and reads \cite{Hung:2011nu}
\be
V_{\Sigma, \rm univ} = \frac{\pi^{d/2}}{\Gamma (d/2)}  \times 
\begin{cases}
    (-1)^{\frac{d}{2}-1} \, \frac{2}{\pi} \log \le \frac{2R}{\delta} \ri & \text{if $d$ even} \\
    (-1)^{\frac{d-1}{2}} & \text{if $d$ odd}
\end{cases}
\ee
We specialize below to two cases: vanishing chemical potential $\mu=0$, and the fine-tuned setting where the solution becomes supersymmetric.

\subsubsection{Uncharged black hole}

When $Q=\mu=0$, the gauge field \eqref{eq:zeroth_order_solution_gauge} vanishes and the expressions for the various thermodynamic quantities simplify.
From now on, we refer to this setting as the \textit{uncharged case}.
For later convenience, it is useful to mention that the critical mass \eqref{eq:critical_mass_radius} becomes
\be
m_{\rm cr} = - \frac{2}{d-2} \le \frac{d-2}{d} \ri^{\frac{d}{2}} \, L^{d-2} <0 \, ,
\label{eq:mcr_nocharge}
\ee
and the mass parameter \eqref{eq:relation_mass_charge} with vanishing charge is directly related to the horizon radius via the identity
\be
m = \frac{r_h^{d-2}}{L^2} (r_h^2 - L^2) \, .
\label{eq:mass_parameter_uncharged}
\ee
Let us now apply the replica trick. 
The grand-potential in eq.~\eqref{eq:Renyi_from_grand_potential} reduces to the free energy $\mathcal{G}(T_0, \mu=0)=F(T_0)$, while the horizon radius and the R\'{e}nyi entropy become
\begin{subequations}
\label{RenyiAndXn}
\be
x_{n} = \frac{1}{d \, n} + \sqrt{\le \frac{1}{d \, n} \ri^2 + \frac{d-2}{d} } \, ,
\label{eq:holo_xn}
\ee
    \be
S_n = \pi V_{\Sigma} \le \frac{L}{\ell_{\rm P}} \ri^{d-1} 
\frac{n}{n-1} \left[ 2- x_n^{d-2} \le 1 + x_n^2 \ri  \right] 
\, . 
\label{eq:holo_Renyi_entropy}
\ee
\end{subequations}
The R\'{e}nyi entropies \eqref{eq:holo_Renyi_entropy} are defined with index $n \in \mathbb{N}$, but can be analytically continued to any real value $n \geq 0$.
Besides the entanglement entropy $\lim_{n\to1}S_n$, two relevant and peculiar cases are the following. 
The first one is the limit $n \rightarrow \infty$, giving
\be
E_0 := S_{\infty} = \lim_{n\to\infty}S_{n}
=  \pi V_{\Sigma} \le \frac{L}{l_P} \ri^{d-1} \mathcal{E}(d) \, ,
\qquad    \
\mathcal{E}(d) := 2-2 \, \frac{d-1}{d-2} \le \frac{d-2}{d} \ri^{\frac{d}{2}} \, , 
\label{eq:def_E0}
\ee
where $E_0 = -\log \lambda_{\rm max}$ denotes the minimal eigenenergy of the modular Hamiltonian $K = -\log \rho_A$, and $\lambda_{\rm max}$ is the largest Schmidt coefficient of the reduced density matrix.
In the previous formula, we denoted with $\mathcal{E}(d)$ a convenient function of the spacetime dimensions $d$, which satisfies $\mathcal{E}(d) \in [1 , 2-\frac{2}{e}]$ for $d \in [2,\infty]$.

The second interesting case is the opposite limit $n \rightarrow 0$, which defines 
\be
S_0 := \lim_{n \rightarrow 0} S_n = \log [\mathcal{R}] \, ,
\ee
where $\mathcal{R}$ denotes the rank of the reduced density matrix \cite{Hung:2011nu}. 
The R\'{e}nyi-0 received a physical interpretation in terms of free energy in reference \cite{Galante:2013wta}, and a relation to the relative entropy in \cite{Agon:2023tdi}. 

Next, we remind the reader that the replica trick defines the horizon radius in terms of the dimensionless quantity \eqref{eq:holo_xn} via $r_h = L x_n$.
Let us study the implications of the range of this quantity on the black hole thermodynamics.
When the replica index is taken to $n=1$, we get $x_1=1$, and the spacetime reduces to Rindler AdS.
In particular, the thermal entropy \eqref{eq:thermal_entropy_BH} reduces to
\begin{equation}
\label{Srindler}
    \sri
    :=  S_{\rm therm}(n=1) =   2\pi V_\Sigma\left(\frac{L}{\ell_P}\right)^{d-1}
    =
    \frac{2E_0}{\cE(d)}\, ,
\end{equation}
where in the last step we compared the result with the minimal eigenenergy of the modular Hamiltonian introduced in eq.~\eqref{eq:def_E0}.
The quantity $x_n$ decreases monotonically with $n$ for $n \geq 0$. It is bounded in the following range 
\begin{equation}
    \sqrt{\frac{d-2}{d}}=: x_\infty\leq x_n \leq x_1=1 \, ,
    \qquad
    n\in[1,\infty)
    \label{eq:range_xn}
\end{equation}
and it exceeds 1 in the range $n\in(0,1)$.
It is worth noticing that the lowest value $x_{\infty}$ corresponds via eq.~\eqref{eq:mass_parameter_uncharged} to the smallest mass parameter $m=m_{\rm cr}$, while the opposite limit $n \rightarrow 0$ implies $m \rightarrow \infty$.
Consequently, the horizon radius always satisfies $r_h \geq L \sqrt{\frac{d-2}{2}} \gg \ell_P$, rendering the classical holographic analysis well-posed.

\subsubsection{Supersymmetric black hole}

The topological black hole in 3+1 dimensions with blackening factor \eqref{eq:blackening_factor_Einstein_Maxwell} can be embedded in $\mathcal{N}=2$ gauged supergravity, resulting in a 1/2 BPS solution once the following condition is imposed \cite{Alonso-Alberca:2000zeh,Nishioka:2014mwa}:
\be
m = 2 i  Q \, . 
\label{susy_rel}
\ee
Importantly, one can show that this condition \textit{holds in general number of dimensions} (for instance, see \cite{Hosseini:2019and} for the application in 5+1 dimensions).
For this reason, in our analysis the phrasing ``SUSY solution" will always imply that eq.~\eqref{susy_rel} is imposed.
Plugging this BPS condition inside eq.~\eqref{eq:critical_mass_radius}, we find that the supersymmetric black hole admits critical radius and mass given by
\be
m_{\rm cr} = - \frac{2 L^{d-2}}{d-1} \le \frac{d-2}{d-1} \ri^{d-2} <0 \, , \qquad 
r_{\rm cr} = L \, \frac{d-2}{d-1} \, .
\label{eq:critical_mass_susy}
\ee
Let us now consider the $n$-fold replicated geometry.
Using the SUSY condition \eqref{susy_rel} inside eqs.~\eqref{eq:relation_mass_charge}, \eqref{eq:chemical_potential_and_charge} and \eqref{eq:definition_xn_charged}, we find that the chemical potential reads
\be
\frac{\mu_{\text{SUSY}} \ell_*}{2 \pi L} = i \sqrt{\frac{2}{(d-1)(d-2)}}  \frac{n-1}{n} \, .
\label{eq:chemical_potential_SUSY}
\ee
Consequently, the mass, charge and the dimensionless horizon radius $x_n$ in the replicated geometry become functions of the replica index $n$ only.
Explicitly, we find
\begin{subequations}
    \be
x_n =   \frac{(d-2)n+1}{(d-1)n} \, ,
\qquad
n=\frac{1}{(d-1)x_n-(d-2)} \, ,
\label{eq:xn_SUSY}
\ee
 \be
S_n = \pi V_{\Sigma} \le \frac{L}{\ell_{\rm P}} \ri^{d-1} 
\frac{n}{n-1}\left[1+x_n(2-x_n-2x_n^{d-2})\right] 
\, , 
\label{eq:SUSY_Renyi_entropy}
\ee
\label{eq:holo_Renyi_entropy_SUSY_sec2}
\end{subequations}
where the latter expression is referred to as the SUSY R\'{e}nyi entropy, obtained by plugging the chemical potential \eqref{eq:chemical_potential_SUSY} inside eq.~\eqref{eq:charged_Renyi_holo}.
First of all, we observe that the expression \eqref{eq:holo_Renyi_entropy_SUSY_sec2} reduces to \eqref{RenyiAndXn} when $d=2$, implying that the two-dimensional case naturally has a supersymmetric enhancement. 
This is a consequence of the charged R\'{e}nyi entropy being independent of the chemical potential in $d=2$, as observed below eq.~\eqref{eq:charged_Renyi_holo}.

Next, it is again useful to consider the limit $n\rightarrow \infty$ of the R\'{e}nyi entropy, which defines the smallest modular energy $E_0 = -\log \lambda_{\rm max}$ as follows:
\be
E_0 :=
S_{\infty} 
=  
\frac{\sri}{2} \, \mathcal{E}_{\rm s}(d) \, ,
\qquad    
\mathcal{E}_{\rm s}(d)
:= 
1 +\frac{(d-2)d}{(d-1)^2} -2 \, \le \frac{d-2}{d-1} \ri^{d-1} \, .
\label{eq:def_E0_SUSY}
\ee
The quantity $x_n$ in eq.~\eqref{eq:xn_SUSY} is a monotonically decreasing function of $n$ for fixed $d$.
In particular, it is bounded in the following range
\be
\frac{d-2}{d-1} = x_{\infty} \leq x_n \leq x_1=1 \, , \qquad
n \in [1, \infty) \, ,
\label{eq:range_xn_susy}
\ee
where $x_1=1$ corresponds to Rindler AdS in the bulk.
In the range $n \in (0,1)$, we have $x_n >1$.
Since the horizon radius is determined by $r_h = L x_n$, we observe that the lowest value $x_{\infty}$ corresponds to the critical radius $r_{\rm cr}$ in eq.~\eqref{eq:critical_mass_susy}.
Therefore the inequality $r_h \geq L \, \frac{d-2}{d-1} \gg \ell_P$ is always satisfied, implying that we can trustfully apply holography to investigate the properties of the dual CFT.

\subsection{Shape deformations of the entangling surface}
\label{ssec:shape_def_review}

The previous subsections focused on the case of a spherical (or planar) entangling surface $\Sigma$, when the conformal mapping in section~\ref{ssec:map_hyperbolic_space} can be performed.
It turns out that the R\'{e}nyi entropies \eqref{eq:def_charged_Renyi} can be still computed using tools from holography, if a small shape deformation of the entangling surface is performed \cite{Dong:2016wcf,Bianchi:2016xvf,Baiguera:2022sao}.\footnote{Previous investigations on the dependence of R\'{e}nyi and entanglement entropy on the shape of the entangling surface, based on CFT techniques, were employed in \cite{Solodukhin:2008dh,Mezei:2014zla,Allais:2014ata,Bianchi:2015liz}.}
In this regard, we remind the reader that the boundary conditions for the replica trick can be implemented by inserting a codimension-two twist operator $\tau_n$ at the location of $\Sigma$ \cite{Hung:2011nu,Hung:2014npa}.
In the presence of a global charge, $\tau_n$ are promoted to dressed twist operators $\tilde{\tau}_n (\mu)$, carrying a non-trivial Dirac sheet which generates an Aharonov-Bohm flux $-i n \mu$ when moving from one copy to the next one in the $n$-fold geometry \cite{Belin:2013uta}.

This framework allows us to reformulate the problem using the tools of defect CFTs (dCFTs).
In the following, we denote the directions orthogonal to the defect by $x^a = \lbrace x^1, x^2 \rbrace,$ and the parallel directions by $y^i,$ with $i \in \lbrace 1, 2 , \dots , d-2 \rbrace$.
The presence of twist operators breaks translational invariance along the directions orthogonal to $\Sigma$, leading to the appearance of a contact term in the Ward identity for the conservation of the energy-momentum tensor
\be
\label{eq:Warddisp}
\partial_{\mu} T^{\mu a}_{\rm tot} (x,y) = \delta_{\Sigma} (x) D^a (y) \, ,
\ee
where $D^a$ is the displacement operator.

For a spherical (or flat) entangling surface, the first non-trivial correlator is the two-point function\footnote{\label{foot:1pt_displacement}
The introduction of a UV cutoff breaks the scale invariance of the CFT and leads to a non-vanishing one-point function of the displacement operator. However, using a regularization scheme that preserves conformal invariance such as dimensional regularization, one can show that all the terms in $\langle D_a (y) \rangle_{n, \mu}$ can be cancelled by the introduction of appropriate counterterms. Deformations which do not change the shape but only increase/decrease the radius of the spherical entangling surface are purely described by this kind of scheme-dependent terms, therefore we will neglect them since they are not universal. We thank Lorenzo Bianchi and Marco Meineri for discussions on this point.}
\be
\langle D_a (y) D_b (y')  \rangle_{n, \mu} = \delta_{ab} \, \frac{C_D (n,\mu)}{(y-y')^{2(d-1)}} \, ,
\label{eq:two_point_function_DD}
\ee
where  $\langle \cdots \rangle_{n,\mu} = \langle \cdots \, \tilde{\tau}_n (\mu) \rangle $  denotes correlation functions evaluated in the presence of the generalized twist operator $\tilde{\tau}_n (\mu).$
In the previous expression, the coefficient $C_D(n,\mu)$, also known as \textit{displacement norm}, is part of the dCFT data, since the normalization of $D_a$ is fixed by the Ward identity \eqref{eq:Warddisp}.
The other relevant dCFT quantity comes from the one-point function of the energy-momentum tensor in the presence of the defect, which reads \cite{Billo:2016cpy}
\be
\langle T_{ij} (z) \rangle_{n,\mu} = - \frac{h_n (\mu)}{2 \pi n} \frac{\delta_{ij}}{|x|^d} \, , \qquad
\langle T_{ab} (z) \rangle_{n,\mu} = \frac{h_n (\mu)}{2 \pi n} \frac{1}{|x|^d}
\le (d-1) \delta_{ab} - d \, \frac{x_a x_b}{x^2}  \ri \, .
\label{eq:one_pt_function_stress_tensor}
\ee
The dCFT parameter $h_n(\mu)$ is referred to as the conformal dimension of the twist operator.

Let us now consider a perpendicular small displacement of the entangling surface $\Sigma$ of the form
\be
\label{deformation}
\delta X^{\mu} = \delta^{\mu}_a f^a \, ,
\ee
where $f^a (y)$ is the profile of the deformation.
The physical response of the system is measured by the variation of the partition function \cite{Bianchi:2015liz}
\be
 \delta \log Z_n (\mu) = \frac{1}{2} \int_{\Sigma} dw \int_{\Sigma} dw' \, 
f^a (w) f^b (w') \langle D_a (w) D_b (w') \rangle_{n,\mu} + \mathcal{O} (f^4) \, .
\label{varpart}
\ee
Applying the identity \eqref{eq:def_charged_Renyi} and using the explicit form \eqref{eq:two_point_function_DD} of the two-point function for $D^a$, we obtain the response of the charged R\'{e}nyi entropies as
\be
 \delta S_n(\mu)  = \frac{n \, C_D(1,\mu) - C_D (n,\mu)}{2(n-1)}  
 \int_{\Sigma} dw \int_{\Sigma} dw' \, 
\frac{f^a (w) f_a (w')}{(w-w')^{2(d-1)}}  + \mathcal{O} (f^4) \, .
\label{eq:variation_Sn_in_terms_of_CD}
\ee
In this work, we restrict to either the case without charge ($Q=\mu=0$), or when the SUSY condition \eqref{eq:chemical_potential_SUSY} is imposed.
In both settings, $C_D$ becomes a function of the replica index $n$ only, and satisfies the property $C_D(n=1)=0$. 

The conformal mapping described in section~\ref{ssec:map_hyperbolic_space} can be appropriately deformed to account for the deformation of $\Sigma$ from its spherical (or planar) shape, and $C_D$ can be numerically computed using holographic renormalization in a deformed black hole background \cite{Dong:2016wcf,Bianchi:2016xvf,Baiguera:2022sao}.
For the purposes of this work, there are two relevant cases where analytic results can be achieved.

\paragraph{Uncharged case.}
When $\mu=0$, the behaviour of $C_D$ in the limit $n\rightarrow 0$ is given by \cite{Bianchi:2016xvf}
\begin{subequations}
    \be
C_D(n) = - \le \frac{1}{d n} \ri^{d-1}  C_T \,  \frac{2^{d-1} \pi^2}{d+1} + \mathcal{O}(n) \, , 
\label{eq:CD_smalln}
\ee
\be
C_T =  2^{d-2} \pi^{- \frac{d+1}{2}} d(d+1) \Gamma \le \frac{d-1}{2} \ri
\le \frac{L}{\ell_{\mathrm{P}}} \ri^{d-1} \, ,
\label{eq:def_CT}
\ee
\label{eq:small_CD}
\end{subequations}
where $C_T$ is the central charge entering the two-point function of the energy-momentum tensor in the absence of a defect (for the normalization of the previous formula, see eq.~(2.27) in \cite{Belin:2013uta}).

\paragraph{Supersymmetric case.}
For supersymmetric theories, it was argued that the generalized twist operators satisfy the following identity\footnote{A holographic proof of this identity was given in \cite{Baiguera:2022sao}. Previously, proofs in $d=3,4$ were derived using CFT techniques in \cite{Bianchi:2019sxz,Lewkowycz:2013laa,Drukker:2019bev,Fiol:2015spa}.  }
\begin{equation}
C_D^{\rm conj} (n,\mu) = d \, \Gamma \le \frac{d+1}{2} \ri \le \frac{2}{\sqrt{\pi}} \ri^{d-1} h_n (\mu)  \, .
\label{eq:conj_CD}
\end{equation}
In a holographic CFT, the conformal weight of the twist operator is known explicitly, \ie
\be
h_n (\mu) = \pi n \le \frac{L}{\ell_{\mathrm{P}}} \ri^{d-1}
\left[  x_n^{d-2} (1- x_n^2) - \frac{d-2}{2(d-1)} \le \frac{\mu \ell_*}{2 \pi L} \ri^2 x_n^{d-2}  \right] \, .
\label{eq:weight_twist_operator}
\ee
In particular, when the BPS condition \eqref{eq:chemical_potential_SUSY} is imposed, it simplifies to
\be
h_n^{\rm SUSY} = \pi n \le \frac{L}{\ell_{\mathrm{P}}} \ri^{d-1}
\left[  x_n^{d-2} (1- x_n^2) + \frac{(n-1)^2}{n^2 (d-1)^2}  x_n^{d-2}  \right] \, , 
\label{eq:hn_susy}
\ee
where $x_n$ was defined in eq.~\eqref{eq:xn_SUSY}.
The conjecture \eqref{eq:conj_CD} will be explicitly used in this work to study the effect of shape deformations in the SUSY case.

%% file: Draft_Sections/HolographicEDOS.tex
\section{The dual to the density of states}
\label{sec:dual_DOS}

We aim to determine the density of states in the entanglement spectrum associated with the reduced density matrix of a holographic CFT.
Section~\ref{ssec:DOS_from_Renyi} is the main core of this work, since it describes the general strategy to determine the density of states from the R\'{e}nyi entropies computed in holography.
We show in section~\ref{ssec:DOS_low_energies} that the density of states contains a term proportional to $\delta(E-E_0)$, where $E_0$ is the lowest eigenvalue of the modular Hamiltonian, plus a non-trivial contribution at energies $E>E_0$.
The latter is then evaluated using a saddle point approximation.
While the methods that we develop in this section are general, we focus for definiteness on the case where the CFT is holographically dual to hyperbolic black holes with vanishing charge.

\subsection{Density of states from the R\'{e}nyi entropy}
\label{ssec:DOS_from_Renyi}

In this subsection we relate the density of states (DOS) to the R\'{e}nyi entropies of a $d$--dimensional holographic CFT. 
Let $\rho_A$ be the reduced density matrix associated with a subregion $A$ delimited by an entangling surface $\Sigma$, and $K=-\log(\rho_A)$ the modular Hamiltonian with eigenvalues $\{E_i\}$. 
In the standard case with vanishing charge $Q=0$, the R\'{e}nyi entropy defined in eq.~\eqref{eq:def_charged_Renyi} simplifies to 
\be
S_{n} = \frac{1}{1-n} \, \log \Tr (\rho_A^n)  \, ,
\ee
where the involved trace can be interpreted as follows in terms of a partition function in the replicated geometry:
\be
 Z_n := \text{Tr}(\rho_A^n) = e^{(1-n)S_n} \, .
\label{eq:partition_function_Zalpha}
\ee
By introducing the DOS in the entanglement spectrum 
\be
D(E) = \sum_i \delta(E-E_i) \, ,
\ee
we can re-express the partition function as
\be
Z_n = \int_0^{\infty} D(E) e^{-n E}dE \, .
\label{eq:partition_function_DOS}
\ee
In other words, $e^{(1-n)S_n}$ is the Laplace transform of $D(E)$, and thus the inverse Laplace transformation can be exploited to find the DOS in the entanglement spectrum
\be 
\label{invLap}
D(E) = \frac{1}{2 \pi i} \int_{\mathcal{C}} e^{n E} e^{(1-n)S_n} dn \, ,
\ee
where now $n \in \mathbb{C}$, and the conventions for the inverse Laplace transform are collected in appendix~\ref{secInvLaplace}.
The integration contour $\mathcal{C}$ runs parallel to the imaginary axis and
it is located on the right of all the singularities of $e^{(1-n)S_n}$.

At this point, it is important to describe the analytic properties of the R\'{e}nyi entropies, that are the input of our calculation.
The entanglement spectrum can contain more information than the set of R\'{e}nyi entropies $S_n$ (with $n \in \mathbb{N}$), unless certain regularity conditions are satisfied \cite{alma99671493502466,BERG199527}. 
The holographic computation automatically gives us an analytic continuation for all $n \in \mathbb{R}$.
After further continuing the replica index $n$ to the complex plane, the inverse Laplace transform allows us to extract a smooth entanglement spectrum for $E>E_0$, where $E_0$ is the minimal eigenenergy.\footnote{Notice that the exact entanglement spectrum is expected to be a discrete sum of Dirac $\delta$-distributions instead of a smooth function \cite{Mukhametzhanov:2019pzy}. Therefore, holography only provides a coarse-grained evaluation of the density of states. We will come back to this issue in section~\ref{sec:Future}.}

\begin{figure}[ht]
    \centering
    \includegraphics[width=0.55\textwidth]{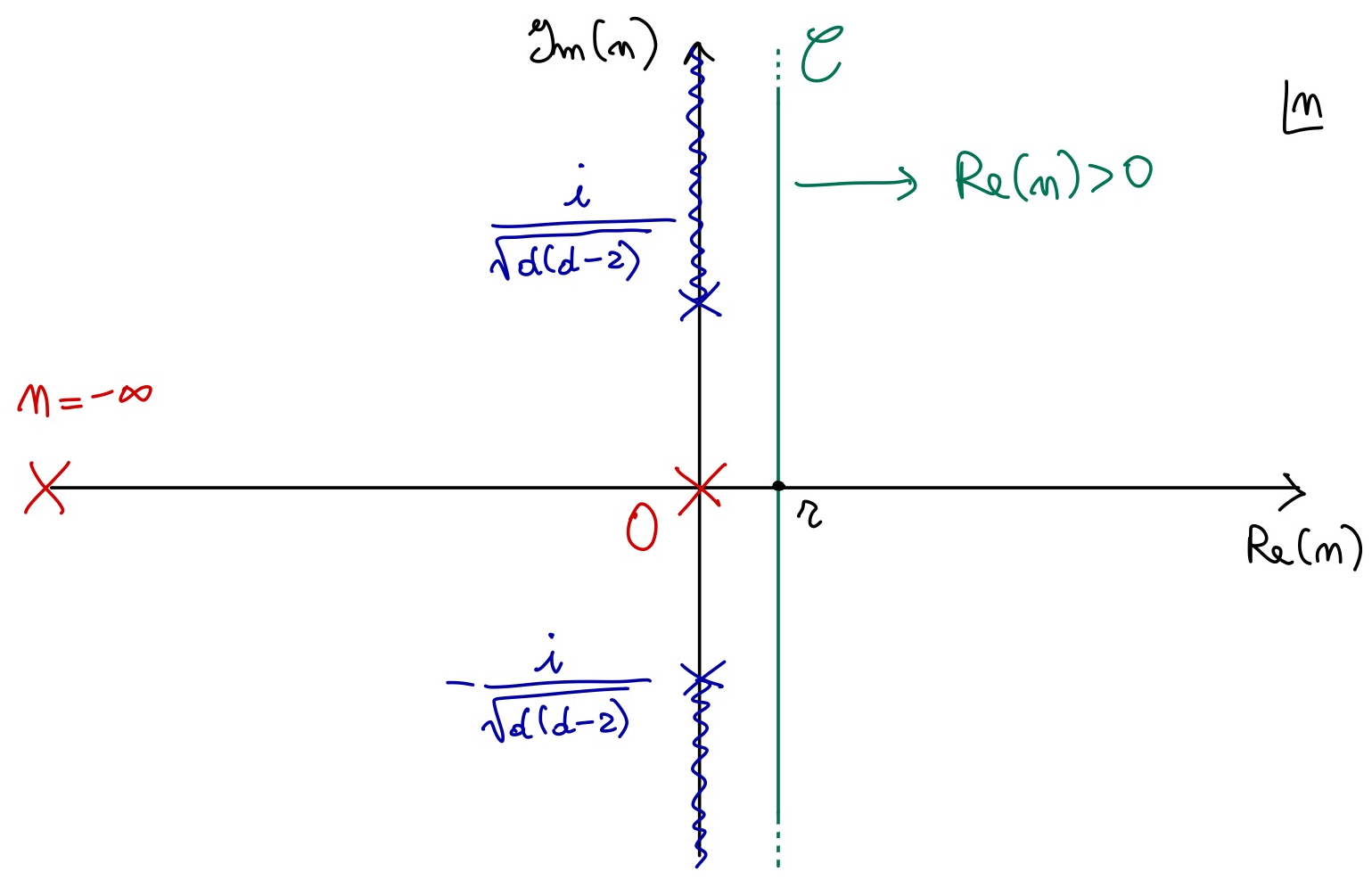}
    \caption{Structure of singularities of $e^{(1-n)S_n}$ for the holographic R\'{e}nyi entropies in eq.~\eqref{eq:holo_Renyi_entropy}. Red crosses denote essential singularities, while the blue wiggled lines are the branch cuts. The integration contour $\mathcal{C}$ for the inverse Laplace transform \eqref{invLap} is depicted in green. }
    \label{fig:singularities_DOS}
\end{figure}

In order to classify the type and location of all the singularities in the complex plane, we need to specify the explicit form of the R\'{e}nyi entropy. 
In the following, we focus on the case of holographic CFTs with vanishing chemical potential ($\mu=0$).
Importing the holographic result \eqref{eq:holo_Renyi_entropy}, we find
\be
\exp \left[ (1-n) S_n \right] = \exp(-\pi V_{\Sigma} \le \frac{L}{\ell_{\rm P}} \ri^{d-1} 
n \left[ 2- x_n^{d-2} \le 1 + x_n^2 \ri  \right] ) \, ,
\label{eq:exp_Renyi_uncharged}
\ee
where $x_n$ was defined in eq.~\eqref{eq:holo_xn}.
Since $x_n$ depends on the R\'{e}nyi index through a square root that vanishes at 
\be
 n = \pm  i \,  \frac{1}{\sqrt{d(d-2)}} \, ,
 \label{eq:branch_points}
\ee
we locate the branch cut along the imaginary axis, connecting these branch points through the point at $\pm i \infty$ in the complex plane.\footnote{
Notice that the branch cut cannot be chosen to run along the imaginary axis, directly connecting the two branch points in eq.~\eqref{eq:branch_points}, because it would encounter the essential singularity at $n=0$.}  
Furthermore, the function \eqref{eq:exp_Renyi_uncharged} also presents essential singularities when the argument of the exponential goes to $\infty$, which only happens for
\be
n = -\infty \quad \vee  \quad 
n = 0 \, ,
\ee
due to the definition of $x_n$.
The structure of singularities of $e^{(1-n)S_n}$ is depicted for convenience in fig.~\ref{fig:singularities_DOS}.

This implies that the inverse Laplace transform is always well-defined in the region $\Re(n)>0$. Hence, the contour $\mathcal{C}$ in eq.~\eqref{invLap} selects a path parallel to the imaginary axis, which intersects the real axis at a strictly positive value $r>0$. We can specify the integral \eqref{invLap} to be
\begin{equation}\label{eq:DOScontour2}
    D(E) 
    =
    \lim_{K\to\infty}\int_{r - i K}^{r + i K} e^{n E} e^{(1-n)S_n}\frac{dn}{2\pi i}
    =
    \lim_{K\to\infty}\int_{r - i K}^{r + i K} e^{f(n)}\frac{dn}{2\pi i}
    \, ,
\end{equation}
where we conveniently repackaged the exponential into a new function $f(n)$. 
Since $S_n$ is proportional to $1/G_N$, we can apply a saddle point approximation to the integrand for sufficiently small $G_N$ (equivalenlty, for large central charge $C_T \gg 1$).\footnote{Using $8 \pi G_N= \ell_P^{d-1}$ (see footnote~\ref{foot:Planck_length}) inside eq.~\eqref{eq:holo_Renyi_entropy}, we get the dependence $S_n \propto 1/G_N$.   } The function $f(n)$ is analytic on the right-half plane, therefore we can Taylor-expand it at some locus $n_*$ with $\Re(n_*)>0$,
\begin{align}
    f(n)
    &:=
    n E+(1-n)S_n\notag\\
    &=
    f(n_*)+f'(n_*)(n-n_*)+\frac{1}{2}f''(n_*)(n-n_*)^2+\dots
    \label{eq:function_fn}
\end{align}
Choosing $n_*$ to be a saddle point imposes
\be
f'(n_*)=0 \quad \Rightarrow \quad
E = - \partial_n \bigl[(1-n)S_n\bigr]\bigl|_{n=n_*} \, .
\label{eq:saddle_point}
\ee
Since the energy is a free parameter entering the density of states $D(E)$, this relation provides a dependence $n_*=n_*(E)$. 
We notice that since $(1-n)S_n = \log Z_n$, the above equations are the usual thermodynamic relations where $E$ is the average energy, and $f(n_*)$ is the thermodynamic entropy.

The regime of validity and the computation of the saddle point approximation (also known as \textit{method of steepest descent}) are summarized in appendix~\ref{ssec:saddle_point}.
Since the function $f(n)$ in eq.~\eqref{eq:DOScontour2} is analytic in the right half-plane, we can always use Cauchy's theorem to make the contour $\mathcal{C}$ running through a saddle $n_*$, such that 
\be
r=\Re(n_*)\, . 
\label{eq:assumption_gamma}
\ee
Whenever several saddle points $n_*^{(j)}$ exist in the region $\Re(n)>0$, one can use the coordinate substitution $n=r+i\beta$ to show that
\be
\begin{aligned}
\label{DOSf}
    D(E) 
    &\approx
    \sum_j
    e^{f(n_*^{(j)})}\lim_{K\to\infty}
    \int_{-K}^{ K}\,
    \exp\left[-\frac{f''(n_*^{(j)})}{2}\bigl(\beta-i(r-n_*)\bigr)^2\right]\frac{d\beta}{2\pi } \\
    &=
    \sum_je^{f(n_*^{(j)})}\frac{1}{\sqrt{2\pi f''(n_*^{(j)})}}
    \, .
\end{aligned}
\ee
We point out that there can be several eligible saddles $n_*^{(j)}$ (\ie with a positive real part), therefore we need to identify the saddle giving the leading contribution. 
A precise analysis of the location of the saddle points and the determination of the dominant one will be performed in the high-energy limit in section~\ref{ssec:dominant_saddles}.
In the following, we analyze the saddle points individually and drop the label $j$ to avoid clutter, reinstating it when necessary. 

The condition for the convergence of the integral \eqref{DOSf} reads
\begin{equation}\label{f''positive}
    f''(n_*)
    =
    \p_n^2\bigl[(1-n)S_n\bigr]\bigl|_{n=n_*}
    >0    \, ,
\end{equation}
where $\p_n^2\bigl[(1-n)S_n\bigr]\bigl|_{n=n_*}$ is the variance in the energy. 
Notice that the exponential term inside eq.~\eqref{DOSf} can be further refined as
\begin{equation}
    f(n_*)
    =
    -n_* \partial_n\left[ (1-n)S_n\right]\bigl|_{n=n_*} + (1-n_*)S_{n_*}
    =
    n_*^2 \partial_n\left[ \frac{(n-1)}{n} S_n\right]\biggl|_{n=n_*} \, .
\end{equation}
Here, we started from the definition of $f(n)$ in \eqref{eq:function_fn} and used the saddle point constraint \eqref{eq:saddle_point} to replace $E$ in the first equality; the second equality is simply a reorganization of the terms. Hence, the density of states is written as
\be
D(E) 
\approx 
\sum_j
\sqrt{\frac{1}{2 \pi\partial^2_n\left[ (1-n)S_n\right]|_{n=n_*^{(j)}}}}
\exp{\bigl(n_*^{(j)}\bigr)^2 \partial_n\left[ \frac{(n-1)}{n} S_n\right]\biggl|_{n=n_*^{(j)}}} \, ,
\label{eq:leading_DOS}
\ee
where $n_*^{(j)}$ is a function of $E$, and we re-instated the summation over all the saddle points in the region $\Re(n)>0$. 

Interestingly, we notice that the latter exponential term coincides with the analogous term in eq.~(4) of \cite{Dong:2016fnf}.
This hints towards a holographic interpretation of the density of states given by
\be
D(E)
\approx
\sqrt{\frac{1}{2 \pi\partial^2_n\left[ (1-n)S_n\right]|_{n=n_*}}} \, e^{\frac{A(\text{cosmic brane}_{n_*})}{4 G_N}} \, ,
\label{eq:DOS_brane}
\ee
where $A(\text{cosmic brane}_{n_*})$ is the area of a bulk, backreacting, codimension-two cosmic brane with tension $T_n = \frac{n-1}{4 n G_N}$, homologous to the entangling region.  

\paragraph{Summary of the procedure.}
For convenience, we conclude this section with a summary of our systematic method, that we will apply in the remainder of this work:
\begin{enumerate}
    \item Take as an input the R\'{e}nyi entropy $S_n$ of a certain physical system, and study the structure of singularities of $e^{(1-n)S_n}$.
    Define the inverse Laplace transform \eqref{invLap} by taking a contour $\mathcal{C}$ parallel to the imaginary axis, and located on the right of all the singularities.
    \item Find the set of saddle points satisfying $f'(n_*)=0$, where $f$ is defined in eq.~\eqref{eq:function_fn}.
    \item Check that the condition \eqref{f''positive} is satisfied by all the saddle points located in the allowed region of the complex plane in the variable $n$.
    \item Plug the saddles inside the integrand and compute the density of states using eq.~\eqref{eq:leading_DOS}.
\end{enumerate}


\subsection{The minimal energy or maximal Schmidt coefficient}
\label{ssec:DOS_low_energies}

We now show that the density of states takes the following form:
\be \label{deltaPlusTheta}
D(E) = \delta(E-E_0) + \Theta (E-E_0) \mathfrak{D}(E),
\ee
where $\Theta$ is the Heaviside step function, $\mathfrak{D}$ is the restriction of $D$ to energies larger then $E_0$, and $E_0$
is the smallest modular eigenenergy, defined in eq.~\eqref{eq:def_E0}.

\begin{figure}[ht]
    \centering
    \subfigure[Case $E<E_0$]{ \label{subfig:int1_delta}  \includegraphics[width=0.5\textwidth]{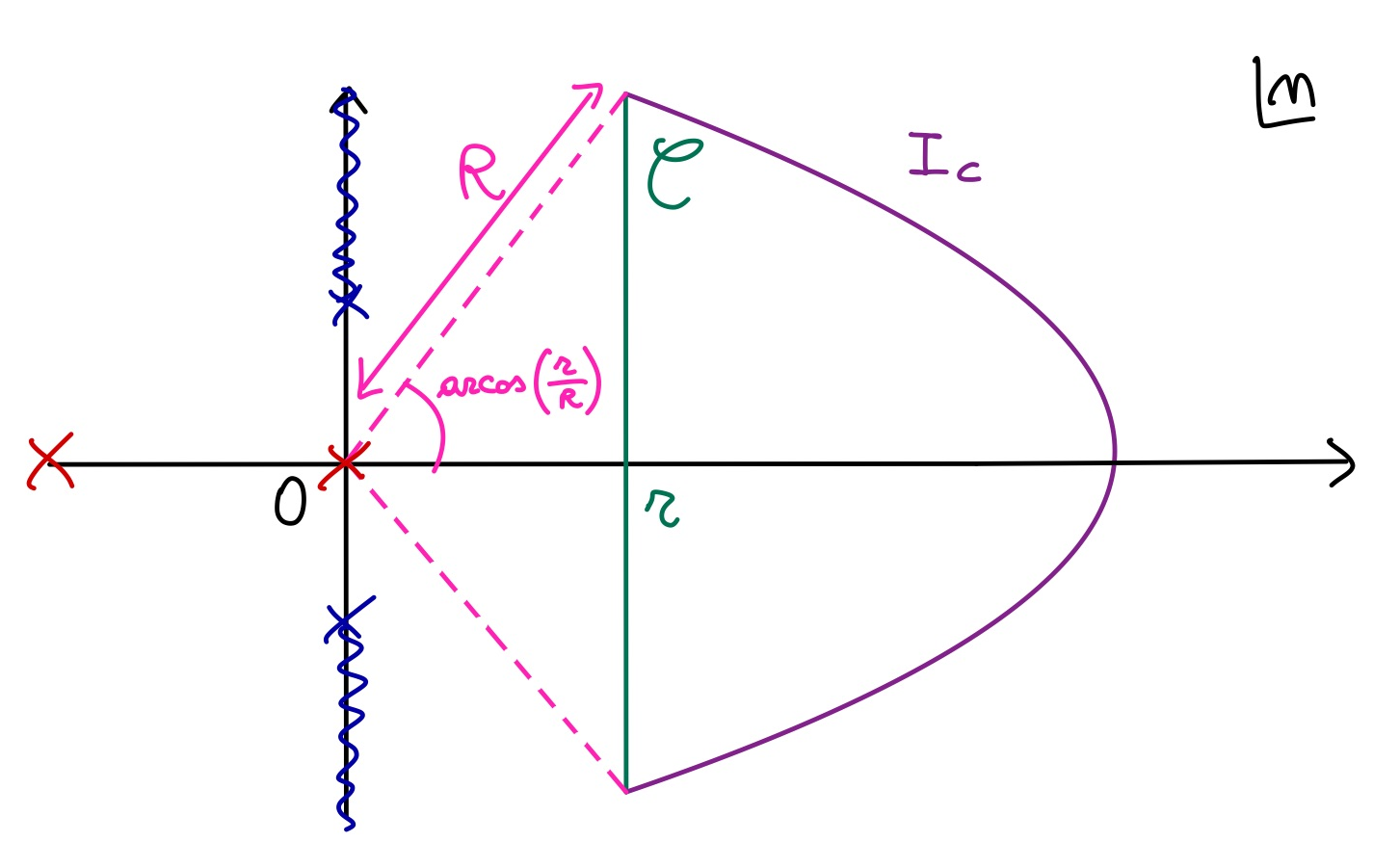} } \quad
    \subfigure[Case $E>E_0$]{ \label{subfig:int2_delta} \includegraphics[width=0.42\textwidth]{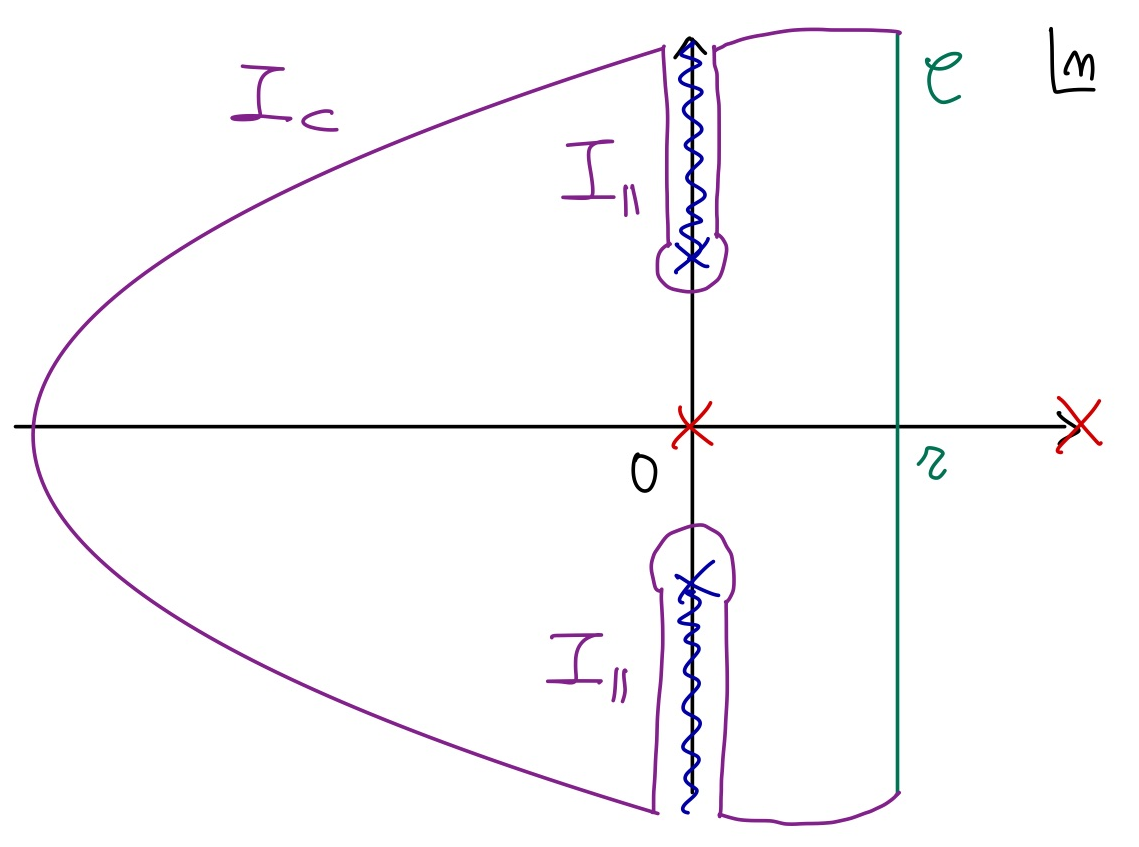}}
    \caption{Closed integration contour used to compute the inverse Laplace transform \eqref{invLap}, composed by the vertical green line $\mathcal{C}$ plus the violet semicircle corresponding to the $I_{\subset}$ contribution.
    The integration region is chosen to avoid the essential singularity at $n= \pm \infty$ (red crosses) of the exponential.
    (a) When $E<E_0$, the semicircle is on the right half of the complex plane. (b) When $E>E_0$, the semicircle is on the left half of the complex plane, and needs to be deformed along $I_{||}$ to avoid the branch cuts.}
    \label{fig:contour_delta}
\end{figure}

To this aim, we introduce the integrals
\begin{equation}
    I = D(E) +I_{\subset} + I_{||} \, ,
    \label{eq:split_integrals_Isub}
\end{equation}
where $D(E)$ is the contribution \eqref{invLap} coming from the vertical green line $\mathcal{C}$ in figure \ref{fig:contour_delta}, $I_{\subset}$ has the same integrand but is evaluated on a semicircle with $|n|=\infty$, $I_{||}$ contains the parts of the integration contour that run along the branch cuts, and $I$ is the full integral over the closed contour composed by the union of the previous ones.
The precise setting is depicted in fig.~\ref{fig:contour_delta}.

The idea is that the integral $I$ can be evaluated by using the residue theorem, but this requires the integrand to only contain isolated singularities inside the closed integration contour.
One can show that the integrand $e^{f(n)}$ in eq.~\eqref{eq:DOScontour2} admits an essential singularity at $n=0$, branch cuts along the imaginary axis starting from the branch points \eqref{eq:branch_points}, and an additional essential singularity located at 
$n=- \infty$ ($n=\infty$) when $E<E_0$ ($E>E_0$).
In order to avoid that the integration region where $I_{\subset}$ is evaluated encounters the essential singularity at infinity, we need to take two different paths. 

In fig.~\ref{subfig:int1_delta} (case $E<E_0$), the violet contour is composed by a semicircle in the right half of the complex plane. The function $e^{f(n)}$ is analytic in such region, and no singularities are encountered.
A trivial application of the residue theorem implies $I=0$.
Since the branch cuts are avoided in this region, we also trivially find $I_{||}=0$.

In fig.~\ref{subfig:int2_delta} (case $E>E_0$), the violet contour is composed by a semicircle on the left half of the complex plane, but this path needs to be deformed to avoid the branch cuts, giving a non-trivial contribution $I_{||} \ne 0$.
The application of the residue theorem inside the closed contour leads to a non-trivial result $I \ne 0$, since an essential singularity at $n=0$ is now included.
We will study the contributions coming from $I-I_{||}$ in the remainder of the paper by using the saddle point approximation.
For the purposes of this section, we denote them by $\Theta(E-E_0) \mathfrak{D}(E)$. 

Now we proceed instead in computing the integral $I_{\subset}$ corresponding to the violet paths in fig.~\ref{fig:contour_delta} with $|n|\to \infty$
\be
\label{eq:DOScontour1theta}
I_{\subset}  = 
\int_{\subset} 
\frac{dn}{2\pi i} \, e^{f(n)} =  \frac{1}{2 \pi}\lim_{R\to\infty} \int_{\frac{\pi}{2}-\frac{r}{R}}^{-\frac{\pi}{2}+\frac{r}{R}}d\theta R e^{i \theta + f(R e^{i \theta})} \, ,
\ee
where $\int_{\subset}$ denotes the semicircle at infinity, and we used the change of variables $n = R e^{i \theta}$.
In the previous step, we already assumed that $r \ll R$ to expand the endpoints of integration as 
\be
\mathrm{arccos} \le \frac{r}{R} \ri = \frac{\pi}{2} - \frac{r}{R} + \mathcal{O} \le \frac{r}{R} \ri^3 \, .
\ee
Using the identity \eqref{eq:split_integrals_Isub} and the definition of $\mathfrak{D}(E)$ introduced above, we obtain
\be
D(E) -\Theta (E-E_0) \mathfrak{D}(E) = - I_{\subset} \, .
\ee
In the limit $R \to \infty$, after plugging in the holographic result \eqref{eq:holo_Renyi_entropy} for the R\'{e}nyi entropy, we then find
\be \label{deltaFunc}
\begin{aligned}
D(E) -\Theta (E-E_0) \mathfrak{D}(E)  & = -  \frac{1}{2 \pi}\lim_{R\to\infty} \int_{\frac{\pi}{2}-\frac{r}{R}}^{-\frac{\pi}{2}+\frac{r}{R}}d\theta R e^{i \theta + R e^{i \theta} (E-E_0)} 
\\
& = - \frac{1}{2 \pi}\lim_{R\to\infty} \int_{\frac{\pi}{2}-\frac{r}{R}}^{-\frac{\pi}{2}+\frac{r}{R}}\frac{d\theta}{i (E-E_0)} \partial_\theta e^{ R e^{i \theta} (E-E_0)} 
\\
& =\lim_{R\to\infty}\frac{\sin(R(E_0-E))}{\pi (E_0-E)} e^{(E-E_0) r} 
\\
& = \delta(E-E_0) e^{(E-E_0)r}  = \delta(E-E_0) \, ,
\end{aligned}
\ee
where we used a representation of the Dirac $\delta$-distribution to move between the last two lines. We note that the essential singularity at $n = \pm \infty$, where the sign depends on the sign of $E-E_0$, is now explicit, and the side to which the contour is closed is chosen accordingly. 
This proves that the density of states takes the form \eqref{deltaPlusTheta}.
A similar statement can be found in eq.~(3.27) of reference \cite{Benjamin:2023qsc}, where a counterterm is included to make the inverse Laplace transform finite.\footnote{The existence of a Dirac $\delta$-distribution in the density of states associated with CFT or entanglement spectra is also discussed, \eg in \cite{Calabrese:2008iby,Behan:2012tc}.}

From now on, we will focus on the evaluation of $\mathfrak{D}(E)$ in eq.~\eqref{deltaPlusTheta} and simply denote it with $D(E)$, unless explicitly specified.

\section{Holographic CFTs dual to uncharged black holes}
\label{sec:holographic_CFT_uncharged}

We apply the method described in section~\ref{ssec:DOS_from_Renyi} to compute the density of states for the vacuum state of a holographic CFT with spherical (or planar) entangling surface. 
We begin in section~\ref{ssec:examples_lowerd} with some analytic examples in lower dimensions.
In section~\ref{ssec:computation_DOS} we move on to an analytic investigation at high energies in general dimensions.
Finally, in section~\ref{ssec:numerical_analysis} we complement our investigations with a numerical approach in several dimensions.

\subsection{Examples}
\label{ssec:examples_lowerd}

As an intuitive illustration of the framework introduced in section~\ref{ssec:DOS_from_Renyi}, we present a detailed calculation of the density of states in dimensions $d=2$ (where an exact result is available in the literature, \eg see \cite{Cardy:1986ie,Calabrese:2008iby,Cardy:2016fqc,Alba:2017bgn}) and $d=4$.

\subsubsection{Two dimensions}
\label{ssec:DOS_2d}

When $d=2$, the DOS is well-known in the CFT literature, since it
can be expressed at high energies in terms of the universal Cardy formula \cite{Cardy:1986ie}, and there is an exact expression valid at all energy scales \cite{Calabrese:2008iby}.
We show here that our approach, outlined in section~\ref{ssec:DOS_from_Renyi}, reproduces the high-energy result.\footnote{In section~\ref{ssec:exact_SUSYDOS_2d3d} we compute the DOS for a supersymmetric-invariant CFT in $d=2,3$ analytically. Since the supersymmetric R\'{e}nyi entropy in two dimensions coincides with the uncharged case, the interested reader can find there an analytic derivation of the result reported in eq.~\eqref{CLformulaBessel} below.}

In $d=2$ the holographic R\'{e}nyi entropies \eqref{eq:holo_Renyi_entropy} simplify to
\be
S_{n}\Big|_{d=2} = \pi V_{\Sigma} \le \frac{L}{\ell_{\rm P}} \ri 
\frac{n+1}{n}  \, ,
\label{eq:2dRenyi}
\ee
where the explicit two-dimensional form $x_{n}=1/n$ in eq.~\eqref{eq:holo_xn} has been used.
The saddle points are identified by the condition \eqref{eq:saddle_point}, which is solved by 
\be
n_*
=
\pm\sqrt{\frac{E_0}{E-E_0}}
\, ,
\label{eq:saddle_points_2d}
\ee
where $E_0$ is the minimal modular energy defined in eq.~\eqref{eq:def_E0}, corresponding to the smallest dual black hole with mass parameter $m=m_{\rm cr}$ in eq.~\eqref{eq:mass_parameter_uncharged}.
Since $E>E_0$, both solutions are real and have opposite signs.
The negative saddle belongs to the region of the complex plane $\mathrm{Re} \, (n)<0$ where the inverse Laplace transform is not defined, and can be thus discarded.
Furthermore the second derivative $f''$ becomes negative on such solution, corroborating its disqualification.

By plugging the positive saddle inside the function $f$ defined in the first line of eq.~\eqref{eq:function_fn}, we find
\be
f''(n_*)
=
2\sqrt{ \frac{\le  E - E_0 \ri^3}{E_0 }}
> 0 \, .
\ee
Therefore, this solution satisfies the convergence requirement \eqref{f''positive}.

Plugging the positive solution $n_*$ inside the DOS \eqref{eq:leading_DOS}, we obtain
\be
D(E)
=
\left[ \frac{E_0}{ (4\pi)^2 \le  E - E_0 \ri^3 } \right]^{\frac{1}{4}} \, 
e^{2\sqrt{E_0 (E - E_0)}}
\, .
\label{eq:intermediate_DOS_2d}
\ee
This expression has been obtained by performing a saddle point approximation of the inverse Laplace transform \eqref{invLap} at leading order in $1/G_N$, with the only assumption that $E>E_0$.

Next, we want to compare the previous result with an approximation of the exact entanglement spectrum obtained in reference \cite{Calabrese:2008iby}.
A series expansion of eq.~\eqref{eq:intermediate_DOS_2d} at large energies $E\gg E_0$ gives
\be
D(E)
\approx
\left[ \frac{E_0}{ (4\pi)^2  E^3 } \right]^{\frac{1}{4}} \, 
e^{2\sqrt{E_0\, E }}
\, ,
\label{eq:DE_2d_mathematica}
\ee
where we only kept the leading-order terms in $E$ in the prefactor and in the exponential.
This equation needs to be compared with the high-energy approximation of the exact DOS, which reads \cite{Calabrese:2008iby}
\be \label{CLformulaBessel}
D(\lambda) = \frac{b \, \theta(\lambda_{\rm max} - \lambda) }{\lambda \sqrt{b \log (\lambda_{\rm max}/\lambda)}} \, I_1 \le 2 \sqrt{b \log \le \frac{\lambda_{\rm max}}{\lambda} \ri } \ri
+ \delta(\lambda_{\rm max} - \lambda)
\, ,
\ee
where $I_1$ is the modified Bessel function of the first kind, $\lambda$ is the distribution of Schmidt coefficients, and $\lambda_{\rm max} = e^{-b}$ its maximum eigenvalue.
After changing variable to $\lambda = e^{-E}$ and expanding at leading order around $E=\infty$, we get
\be   
D(E)|_{\text{ref \cite{Calabrese:2008iby}}} \approx  \le  \frac{b}{(4\pi)^2 E^3} \ri^{\frac{1}{4}}
\, e^{2 \sqrt{b E}}  \, .
\label{eq:Calabrese_Lefevre_approx}
\ee
Comparing with eq.~\eqref{eq:DE_2d_mathematica}, we notice that the two results agree after identifying $b=E_0$, as expected. 
This provides a first non-trivial check of the general procedure developed in section~\ref{ssec:DOS_from_Renyi}. \textit{En passant}, we observe that one might as have expanded \eqref{CLformulaBessel} for large $b$ while keeping $\lambda_{\textrm{max}}/\lambda$ fixed (equivalently keeping $E-E_0$ fixed) to obtain the same leading scaling as \eqref{eq:Calabrese_Lefevre_approx}.

\subsubsection{Four dimensions}

When $d=4$, the equation $f'(n)=0$ admits four roots.
Two of them have negative real part, therefore they are discarded because the inverse Laplace transform is only defined in the region $\Re(n)>0$.
The other two solutions can have positive real part. Explicitly, they are given by
\be
n_*^{(\pm)} 
=
\frac{1}{\sqrt{10}} 
\frac{
\sqrt{ 12 E_0  \le E - E_0  \ri \pm \sqrt{240 E_0  \le  E - E_0  \ri^3 }  } 
}{ E - E_0  }\, \label{d4saddle}.
\ee
Since $E>E_0$, we notice that $n_*^{(+)}$ is always positive and real.
In order to have manageable expressions to manipulate, we will work from now on in the high-energy regime $E \gg E_0$

Evaluating the function $f$ and its second derivative at $n_*^{(+)}$ and keeping only the leading contribution at large $E$, we obtain
\be
f(n_*^{(+)}) \approx 2\le  \frac{4 E_0  (E-E_0)^3}{135} \ri^{\frac{1}{4}} \, ,
\qquad
f''(n_*^{(+)}) \approx 8 \, \le  \frac{5(E-E_0)^5}{12E_0}  \ri^{\frac{1}{4}} > 0 \, ,
\label{eq:f_f2_4d}
\ee
which lead to the following DOS:
\be
D(E)
\approx
\frac{1}{\sqrt{16 \pi}} \le  \frac{12E_0}{5(E-E_0)^5}  \ri^{\frac{1}{8}}  \exp \le  \frac{64 E_0  (E-E_0)^3}{135} \ri^{\frac{1}{4}} \, .
\ee
This expression fits with the general form \eqref{eq:leading_DOS_largeE}, that will be systematically obtained in section~\ref{ssec:computation_DOS} for any number of dimensions $d$ in the high-energy limit.

Finally, we point out that the other candidate saddle $n_*^{(-)}$ is purely imaginary as long as $\frac{E-E_0}{E_0}>\frac{4}{15}$.
This inequality is certainly satisfied in the high energy limit under consideration. Therefore we discard the saddle $n_*^{(-)}$ because its real part vanishes, moving this solution outside of the integration domain of the inverse Laplace transform.

\subsection{Entanglement spectra at large energies in general dimension}
\label{ssec:computation_DOS}

After the experience gained from the previous examples, we are in a position to compute the density of states for general dimension $d$ using the holographic R\'{e}nyi entropies \eqref{eq:holo_Renyi_entropy} as an input. 
In the following, we derive a general solution for the DOS at 
large energies.
An explicit analysis of the saddle points in terms of the replica index $n$ is made harder by the presence of square roots, see for instance the solutions \eqref{d4saddle}.
We circumvent this issue by working instead with the dimensionless event horizon radius $x_n$ in eq.~\eqref{eq:holo_xn}, using the defining relation 
\begin{equation}
\label{definingXn}
    0=dx_n^2-\frac{2}{n}x_n-(d-2) \, .
\end{equation}
The quantity $x_n$ monotonically decreases for positive and real $n$, as reported in eq.~\eqref{eq:range_xn}. When $n\in(0,1)$, $x_n$ can exceed 1, which is crucial for our derivation below. 
Let us first rewrite $f'$ in terms of $x_n$, beginning with eqs.~\eqref{eq:function_fn} and \eqref{eq:holo_Renyi_entropy} to find
\be
\begin{aligned}
    f'(n)
    &=
    E-\sri\left[1-\frac{1}{2}(1+x_n^2)x_n^{d-2}+\frac{x_n^{d-1}}{n}\right] \\
    &=
    E-\sri\left[1+\frac{(d-1)}{2}(x_n^d-x_n^{d-2})\right] \, ,
\end{aligned}
\ee
where we have employed the definition \eqref{Srindler} of the thermal entropy of Rindler AdS. 
In going from the first to the second line, we have used the defining relation \eqref{definingXn} to replace $x_n/n$, thereby completely eliminating the R\'{e}nyi index $n$ from $f'$.
The saddle point condition $f'(n_*)=0$ thus reads
\begin{equation}
\label{saddlesX}
    \frac{2}{d-1}\left(\frac{E}{\sri}-1\right)
    =
    x_{n_*}^d-x_{n_*}^{d-2} \, .
\end{equation}
We observe that the right-hand side of this equality can be expressed in terms of the mass parameter $m$ of the dual (uncharged) hyperbolic black hole presented in eq.~\eqref{eq:mass_parameter_uncharged} (recall that $x_n=r_h/L$). For the left-hand side, this implies
\begin{equation}\label{energyDensityA}
   \frac{m(E)}{L^{d-2}}
   =
   \frac{2}{d-1}\left(\frac{E}{\sri}-1\right)
   =
   \frac{\cE(d)}{d-1}\frac{E-E_0}{E_0}+\frac{m_{\rm cr}}{L^{d-2}}
   \,\in \mathbb{R} \, ,
\end{equation} 
where eqs.~\eqref{eq:mcr_nocharge} and \eqref{eq:def_E0} were employed. This identity is nothing but the relation between the modular energy $E$ of the CFT subregion state and the mass of its dual holographic black hole. 
It is useful to rewrite the saddle point condition \eqref{saddlesX} as
\begin{equation}
    x_{n_*}^{d-2}
    =
    \frac{m(E)L^{2-d}}{x_{n_*}^2-1}
    \quad 
    \text{for}
    \quad
    x_{n_*}\neq1  \, ,
    \quad 
    \text{\textit{i.e.},}
    \quad 
    n_*\neq1 \, .
\end{equation}
This equation allows us to replace powers of $x^{d-2}$ by $x^2$, thereby drastically simplifying polynomials. 
Plugging this identity into the desired function \eqref{eq:function_fn} yields
\begin{align}
    f(n)
    &=
    n\left[E-\sri\left(1-\frac{1}{2}(1+x_n^2)x_n^{d-2}\right)\right]\\
    \Rightarrow 
    f(n_*)
    &=
    \underbrace{\frac{n_*}{2}\frac{m(E)}{L^{d-2}}\,\sri}_{\frac{n_*}{d-1}(E-\sri)}\left(d-1+\frac{x_{n_*}^2+1}{x_{n_*}^2-1}\right)\label{fSaddle} \, .
\end{align}
The second derivative provides constraints on $n_*$ (or equivalently on $x_{n_*}$), 
\begin{align}
    f''(n)
    &=
    \frac{\sri(d-1)}{4}(x_n^2d-(d-2))^3\frac{x_n^{d-3}}{x_n^2d+d-2}\\
    \Rightarrow 
    f''(n_*)
    &=
    \sri\frac{d-1}{4}\frac{m(E)L^{2-d}}{x_{n_*}(x_{n_*}^2-1)}\frac{\left(d(x_{n*}^2-1)+2\right)^3}{d(x_{n*}^2+1)-2}
    >
    0 \, .
    \label{f''saddle}
\end{align}

\subsubsection*{Asymptotic behavior for large energies}

Our analysis so far is valid for any value of the modular energy $E > E_0$.
In the following we restrict to large energies, \ie
\begin{equation}
    \frac{E-E_0}{E_0}
    \gg 1 \, .
    \label{eq:assumptions_cutoff}
\end{equation}
This implies $m(E)\gg L^{d-2}$ and, in particular, we can drop the negative contribution $m_{\rm cr}/L^{d-2}$ in \eqref{energyDensityA}.
Equivalently, this regime can be understood from the bulk side as the case of large, stable black holes with horizon radius $r_h \gg L$.
Therefore, we conclude that the assumption \eqref{eq:assumptions_cutoff} lies inside the range of parameters where holography can be trustfully applied. 

If the left-hand side of eq.~\eqref{saddlesX} is a very large and positive real number, then $x_{n_*}^d$ must be very large itself.
At leading order, this means that the contribution from $x_{n_*}^{d-2}$ is suppressed, reducing the saddle point condition to
\begin{equation}
    x_{n_*}^d=\frac{m(E)}{L^{d-2}}\gg1 \, .
\end{equation}
This equation has $d$ solutions distributed uniformly on a circle in the complex plane
\begin{equation}
     x_{n_*}^{(k)}
     =
     e^{2\pi ik/d}\le\frac{m(E)}{L^{d-2}} \ri^{1/d}\,,
    \quad
    (k=0,1,\dots  d-1)
    \label{eq:xnstar_solution_largeE}
\end{equation}
Working in the regime \eqref{eq:assumptions_cutoff}, the second derivative \eqref{f''saddle} simplifies to
\begin{equation}
    f''(n_*^{(k)})
    =
    \sri\frac{d-1}{4}d^2\le\frac{m(E)}{L^{d-2}} \ri^{1+1/d}\,e^{2\pi i k/d}
    \approx
    \frac{d^2}{2}\le \frac{\cE(d)(E-E_0)^{d+1}}{(d-1)E_0}\ri^{\frac{1}{d}}\,e^{2\pi i k/d}\, .
\end{equation}
where \eqref{eq:def_E0}, \eqref{Srindler} and the leading part of eq.~\eqref{energyDensityA} were used. This quantity must have positive real part for the saddle point approximation to be valid. Hence, we restrict to solutions $n_*^{(k)}$ satisfying 
\begin{equation}\label{maxConstraint}
    \cos(2\pi  k/d)>0\,.
\end{equation}
This condition similarly follows from the range of validity of the inverse Laplace transformation, which requires $\Re(n)>0$. 
To see this, we plug the solution \eqref{eq:xnstar_solution_largeE} inside the identity~\eqref{definingXn} to obtain
\begin{equation}
    n_*^{(k)}
    =
    \frac{2x_{n_*}^{(k)}}{d\,\left(x_{n_*}^{(k)}\right)^2-(d-2)}
    \approx
    \frac{2}{d\,x_{n_*}^{(k)}}
    =
    \frac{2}{d}\le \frac{m(E)}{L^{d-2}} \ri^{-1/d}e^{-2\pi ik/d} \, ,
    \label{eq:nstar_highE_general}
\end{equation}
from which we notice that the constraint $\Re(n_*^{k})>0$ yields \eqref{maxConstraint} once more. 
It should also be noticed that in the high-energy limit, the saddle points $n_*^{(k)}$ hover dangerously close to $n=0$, but one can check that they never reach the imaginary axis. 
Moreover, the saddle points move further away from the line $\mathrm{Re} \, (n)=0$ with increasing dimension $d$. 

At this point we need to comment on the possibility of phase transitions. Such may appear for instance in a CFT admitting a scalar operator with sufficiently low scaling dimension $\Delta$. In this case the dual asymptotically AdS black hole can be unstable at low temperature where a configuration with scalar hair becomes favorable \cite{Dias:2010ma,Hartnoll:2008kx,Hartnoll:2008vx,Gubser:2008zu}. 
This leads to a phase transition of the R\'{e}nyi entropy at a critical value of the replica index $n$ \cite{Belin:2013dva}. Our starting point \eqref{eq:holo_Renyi_entropy} assumes the absence of phase transitions. However, the saddle \eqref{eq:nstar_highE_general} is small, thus residing at high temperatures, where phase transitions are avoided. Thus we believe our results to also be useful for theories with phase transitions at low temperatures.

Plugging the solution \eqref{eq:nstar_highE_general} into eq.~\eqref{fSaddle}, we obtain 
\be
\begin{aligned}
   f(n_*^{(k)})
    \approx
    \frac{d\,n_*^{(k)}}{2}\,\le \frac{m(E)}{L^{d-2}}\ri\,\sri 
    &=
    2\le \frac{E_0}{\cE(d)}\ri^{\frac{1}{d}}\left(\frac{E-E_0}{d-1}\right)^{\frac{d-1}{d}}\,
    e^{-2\pi ik/d} \, .
\end{aligned}
\ee
The real solution $n_*^{(0)}$ clearly satisfies \eqref{maxConstraint}, and has  the favorable property to dominate the sum over saddles, as we argue in appendix~\ref{ssec:dominant_saddles}.
It can thus be plugged straightforwardly into \eqref{DOSf}, while the sum over $j$ can be dropped. We obtain for the contribution stemming from $n_*^{(0)}$
\begin{equation}
    D(E)
    \approx
     \frac{1}{d \sqrt{\pi}}\left(\frac{\cE(d)(E-E_0)^{d+1}}{(d-1)E_0}\right)^{-\frac{1}{2d}} 
    \exp[2\le \frac{E_0}{\cE(d)}\ri^{\frac{1}{d}}\left(\frac{E-E_0}{d-1}\right)^{\frac{d-1}{d}}]\, .
    \label{eq:leading_DOS_largeE}
\end{equation}
This formula provides an explicit expression for the high-energy limit of the DOS $D(E)$  entering eq.~\eqref{deltaPlusTheta} -- recall that we are referring to $\mathfrak{D}$ as $D$. Let us note that the result \eqref{CardyDeformation} was anticipated, yet not derived in \cite{Hung:2011nu}. In particular, the scaling was reported correctly, but the exact form remained unknown.

The micro-canonical entropy is read off to be 
\begin{equation}\label{CardyDeformation}
  \boxed{ 
  S
  =
  \log D(E)
  =
  2\le \frac{E_0}{\cE(d)}\ri^{\frac{1}{d}}\left(\frac{E-E_0}{d-1}\right)^{\frac{d-1}{d}} +\dots
  }
\end{equation}
where the subtraction by $E_0$ is proportional to the central charge \eqref{eq:def_CT} of the theory  
\begin{equation}\label{CasimirEnergy}
    E_0=\frac{V_{\Sigma} \pi^{\frac{d+3}{2}} \mathcal{E}(d)}{2^{d-2}d(d+1)\Gamma\le\frac{d-1}{2}\ri} \, C_T \, .
\end{equation}
It is worth mentioning that $E_0$ is proportional to the volume of the hyperbolic space $V_{\Sigma}$, which is generally scheme-dependent.
If we aim to get physical information from $E_0$, we need to introduce counterterms such that only the universal part of the volume is retained. 
As mentioned in the introduction, we recover the scaling \eqref{eq:Ooguri_intro} and show moreover precisely in which form the entanglement divergences appear, namely in the volume $V_\Sigma$ and thus the minimal modular energy $E_0$. 

The result \eqref{CardyDeformation} coincides with the Cardy formula in two dimensions, as we already verified in section~\ref{ssec:DOS_2d}.
In higher dimensions, it provides a generalization of the Cardy formula for the entanglement spectrum at large $E$ which satisfies the following consistency checks:
\begin{itemize}
    \item It is consistent with the expectations of the thermodynamic limit, where entropy and energy scale as
    \be
S \sim V T^{d-1} \, , \qquad
E \sim V T^{d} \, , 
\ee
as one can determine from the fact that entropy and energy are extensive in the volume $V$, and fixing the powers in the temperature by dimensional analysis.
\item A similar argument as the previous bullet point can be presented for gapped theories at large $N$ (\eg see eqs.~(2.36)--(2.40) of reference \cite{Harlow:2018fse}) to find the scaling of the Bekenstein-Hawking entropy of a dual black hole with compact spatial slices as
\be
S_{\rm therm} = \frac{A}{4 G_N} \approx G_N^{d-1} \, M^{\frac{d-1}{d}} \, .
\ee
By identifying the energy $E$ in the density of states with the mass of the black hole solution, we find the same scaling.
\item It coincides with Verlinde's formula \cite{Verlinde:2000wg}
\be
S \sim \sqrt{E_C (2E-E_C)} \, ,
\ee
once the Casimir energy is expressed in terms of the entropy of the system.  
This can be seen by using the identities
\be
E = E_E + \frac{1}{2} E_C \, , \qquad
E_E \sim S^{\frac{d+1}{d}} \, , \qquad
E_C \sim S^{\frac{d-1}{d}} \, ,
\ee
where $E_E, E_C$ are the extensive and subextensive parts of the energy, respectively.
\item It agrees with the leading contribution to the microcanonical entropy of free CFTs computed in \cite{Behan:2012tc}.
\item It is consistent with a similar formula obtained for CFT spectra in reference \cite{Shaghoulian:2015kta} by using modular forms, and later extended to holography via the computation of the Bekenstein-Hawking entropy \cite{Shaghoulian:2015lcn} (the original holographic interpretation in three bulk dimensions was found in \cite{Strominger:1997eq}).

\item The scaling agrees with eq.~(3.39) of \cite{Benjamin:2023qsc}, where  the density of states of CFT spectra was computed from the partition function on $S^1 \times S^{d-1}$ by means of a thermal effective action.
In the holographic case where their dictionary (4.7) is employed, the prefactor in eq.~\eqref{CardyDeformation} matches with their result, up to normalization of the modular energies by a factor of $2\pi$ and scheme-dependent terms.
While their techniques are valid for arbitrary CFTs, our results only apply to holographic ones; however, we also provide the subleading contribution to the density of states, given by the prefactor in front of the exponential in \eqref{eq:leading_DOS_largeE}.
\end{itemize}

It is important to stress that most of the above-mentioned consistency checks refer to computations of high-energy CFT spectra, while the results presented in this paper involve the study of entanglement spectra (\ie when a bipartition of the physical system is considered).
However, at leading order in the energy, these two computations agree, thus making the generalization of the Cardy formula a universal statement.
A difference between CFT and entanglement spectra should be detected by the
contributions coming from the boundaries of the entangling surface \cite{Ohmori:2014eia,Alba:2017bgn}.
We leave this topic for future studies.


\subsection{Numerical results}
\label{ssec:numerical_analysis}

While the above computations provided us with analytic solutions for the saddle points and for the large-energy limit of the DOS, another possibility to make further steps is to perform a numerical analysis.
In particular, we will analyze the density of states in dimensions $2 \leq d \leq 6$.

We briefly outline the main steps involved in this procedure:
\begin{enumerate}
    \item Fix the dimension $d$ and find the list of candidate saddle points as the numerical solutions to the condition $f'(n_*)=0$.
    \item Pick the dominant contribution to $f(n_*)$ and check that it satisfies the convergence condition for the inverse Laplace transform, \ie $f''(n_*)>0$.
    \item Plug such preferred solution inside $f$ and $f''$ defined in eq.~\eqref{eq:function_fn} and plot the results.
\end{enumerate}

\begin{figure}[ht]
    \centering
    \subfigure[]{  \includegraphics[scale=0.69]{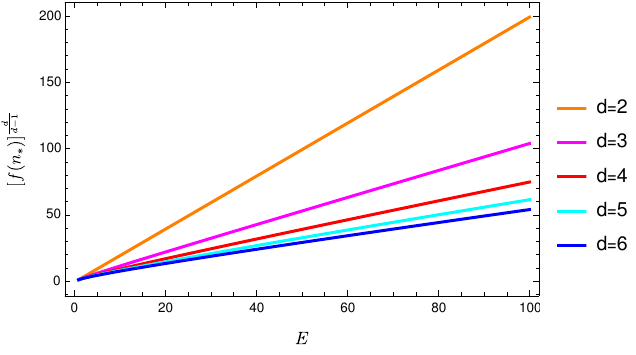}  }
    \subfigure[]{  \includegraphics[scale=0.69]{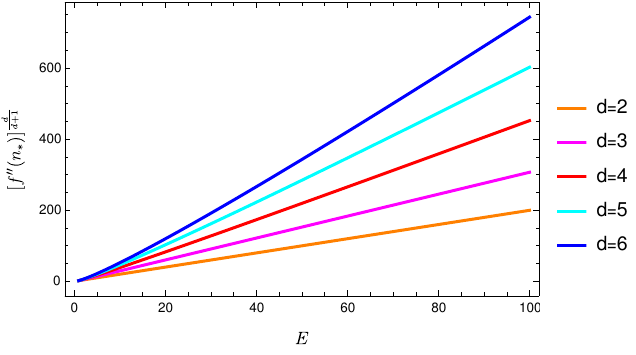}  }
    \caption{Plot of $f^{\frac{d}{d-1}}$ and $(f'')^{\frac{d}{d+1}}$ as functions of the energy $E$, for fixed $\sri=1$. The parameter $d$ refers to the spacetime dimensions of the CFT.}
    \label{fig:plotf_f2}
\end{figure}

The values of $f(n_*)$ and $f''(n_*)$ at the dominant saddle point control the leading and next-to-leading contributions to the density of states.
Following the previous steps, we were able to infer that their leading scaling at high energy as functions of the modular energy for various dimensions read
\be
f(n_*) \sim E^{\frac{d-1}{d}} \, , \qquad
f''(n_*) \sim E^{\frac{d+1}{d}} \, .
\ee
Indeed, the plots collected in fig.~\ref{fig:plotf_f2} confirm this trend because we show that the inverse functions $f^{\frac{d}{d-1}}$ and $(f'')^{\frac{d}{d+1}}$ approach a linear behaviour for large energies.
This result is consistent with the analytic formula \eqref{eq:leading_DOS_largeE} derived in section~\ref{ssec:computation_DOS}.

%% file: Draft_Sections/Susy_DOS.tex
\section{Holographic supersymmetric CFTs}
\label{subsec:DOS_SUSY}

We apply the machinery developed in section~\ref{ssec:DOS_from_Renyi} to a supersymmetric-invariant CFT with holographic dual.
The SUSY case is achieved by the BPS condition~\eqref{susy_rel} in the holographic setting, which requires to consider a $\mathrm{U}(1)$ R-symmetry group on the field theory side.
While we started with two independent parameters $(n, \mu)$, supersymmetry imposes the relation \eqref{eq:chemical_potential_SUSY} between them, rendering the R\'{e}nyi entropy dependent only on the replica index, see eq.~\eqref{eq:holo_Renyi_entropy_SUSY_sec2}. By using the definition \eqref{Srindler} for the thermal entropy of Rindler AdS, we get
\be
S_n^{\rm SUSY} = \frac{\sri}{2} 
\frac{n}{n-1}\left[1+x_n(2-x_n-2x_n^{d-2})\right]
\, ,
\label{eq:holo_Renyi_entropy_SUSY}
\ee
that is considered as the holographic input for the remainder of this section.

We begin by clarifying in section~\ref{ssec:which_DOS_SUSY} which kind of DOS is computed in the SUSY case. 
We then use the simple analytic structure of the SUSY R\'{e}nyi entropy (which is a consequence of the simple form without branch cuts of $x_n$ in eq.~\eqref{eq:xn_SUSY}) to compute the exact DOS in two and three dimensions, see section~\ref{ssec:exact_SUSYDOS_2d3d}.
We proceed in section~\ref{ssec:SUSY_largeE_anyd} with an analytic computation at large energies in any dimension, and we conclude with some numerical examples in subsection~\ref{ssec:SUSY_numerical}.

\subsection{Which density of states?}
\label{ssec:which_DOS_SUSY}

Let us explain to which entanglement spectrum the DOS computed in the SUSY case belongs to.
In the presence of a non-vanishing chemical potential, the partition function $Z_n (\mu)$ reads
\be
Z_n (\mu) = \Tr \left[ \rho_A e^{\mu Q_A} \right]^n \, ,
\ee
where the prescription in eq.~\eqref{eq:general_replica_trick} has been used.
Despite the existence of a non-trivial charge $Q_A$, the reduced density matrix is still associated with a modular Hamiltonian via $\rho_A = e^{-K}$.
Therefore, eq.~\eqref{eq:partition_function_DOS} generalizes to
\be
Z_n (\mu) = \int dE \int dQ \, D(E, Q) e^{-n E+ n \mu Q} \, ,
\label{eq:partition_function_nmu}
\ee
where the density of states $D(E,Q)$ now depends on both the energy and charge, since the reduced density matrix acts on sectors with different values of the conjugate variables $(n, \mu)$.
Here and in the remainder of the work, we drop the subscript referring to the subsystem $A$ on the global charge. 

Let us now assume that the chemical potential and the R\'{e}nyi index are related by the constraint
\be \label{eq:susy-chem}
\mu = \gamma \, \frac{n-1}{n} \, ,
\ee
for a certain constant $\gamma$.
When $\gamma = \frac{2\pi i L}{\ell_*} \sqrt{\frac{2}{(d-1)(d-2)}}$, this identity defines the chemical potential \eqref{eq:chemical_potential_SUSY} in the SUSY case. We reiterate here that this chemical potential corresponds to the supersymmetric point at which the solution of Einstein-Maxwell can be embedded in a supergravity theory.
Since the chemical potential is not arbitrary, we cannot access the full density of state as function of both charge and energy, but only to an integrated version. More precisely, plugging the relation \eqref{eq:susy-chem} in the partition function \eqref{eq:partition_function_nmu} gives
\be
\begin{aligned}
Z_n (\mu_{\rm SUSY}) & = \int d\tilde{E} \int dQ \, D(\tilde{E}+ \gamma Q, Q)  e^{-\gamma Q} e^{-n \tilde{E}} \\
& = \int_0^{\infty} d \tilde{E} \, \tilde{D} (\tilde{E}) e^{-n \tilde{E}} \, ,
\end{aligned}
\ee
where in the first step we performed the change of variables $\tilde{E}= E - \gamma Q$, and in the last step we introduced the integrated DOS:
\be
\tilde{D} (\tilde{E}) \equiv \int dQ \, D(\tilde{E}+ \gamma Q, Q) e^{-\gamma Q} \, . 
\label{eq:Dgamma}
\ee
At this point, one can use the SUSY R\'{e}nyi entropy in eq.~\eqref{eq:holo_Renyi_entropy_SUSY} to write the partition function as
\be
Z_n (\mu_{\rm SUSY}) = e^{(1-n) S_n^{\rm SUSY}} \, ,
\ee
and proceed in the same way outlined in section \ref{ssec:DOS_from_Renyi} to compute the density of states $\tilde{D}(\tilde{E})$ as an inverse Laplace transform.
We obtain
\be
\tilde{D} (\tilde{E})  = \frac{1}{2\pi i} \int_{\mathcal{C}} dn \, e^{n \tilde{E}} e^{(1-n) S_n^{\rm SUSY}} \, ,
\label{eq:DOS_tilde}
\ee
where the integration contour is a vertical line in the complex plane, intersecting the real axis at a value $\mathrm{Re}\, (n) >0$.
The quantity $e^{(1-n) S_n^{\rm SUSY}}$ only admits two essential singularities at $n=-\infty$ and $n=0$.
Contrarily to the uncharged case discussed below eq.~\eqref{eq:exp_Renyi_uncharged}, there are no branch cuts due to the rational structure in $n$ of $x_n$ in eq.~\eqref{eq:xn_SUSY}.

From now on, the tildes on $\tilde E$ and $\tilde D$ are omitted yet always implied to avoid clutter when dealing with the supersymmetric case.

\subsection{Analytic results at arbitrary energy in 2d and 3d}
\label{ssec:exact_SUSYDOS_2d3d}

\begin{figure}[ht]
    \centering
    \subfigure[]{ \label{fig:plotintsusy} \includegraphics[scale=0.17]{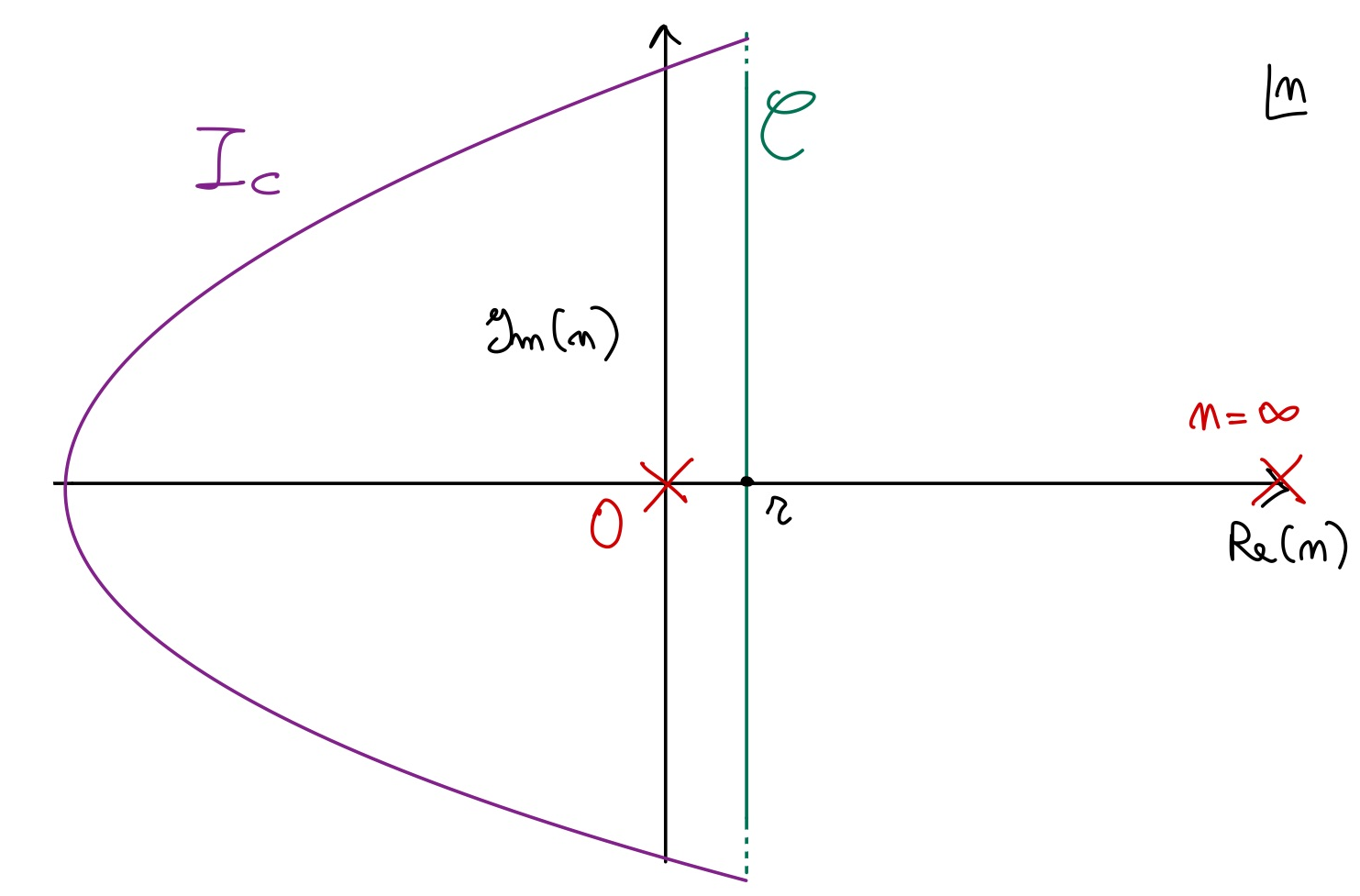}}
    \subfigure[]{\label{fig:deformed_contour} \includegraphics[scale=0.17]{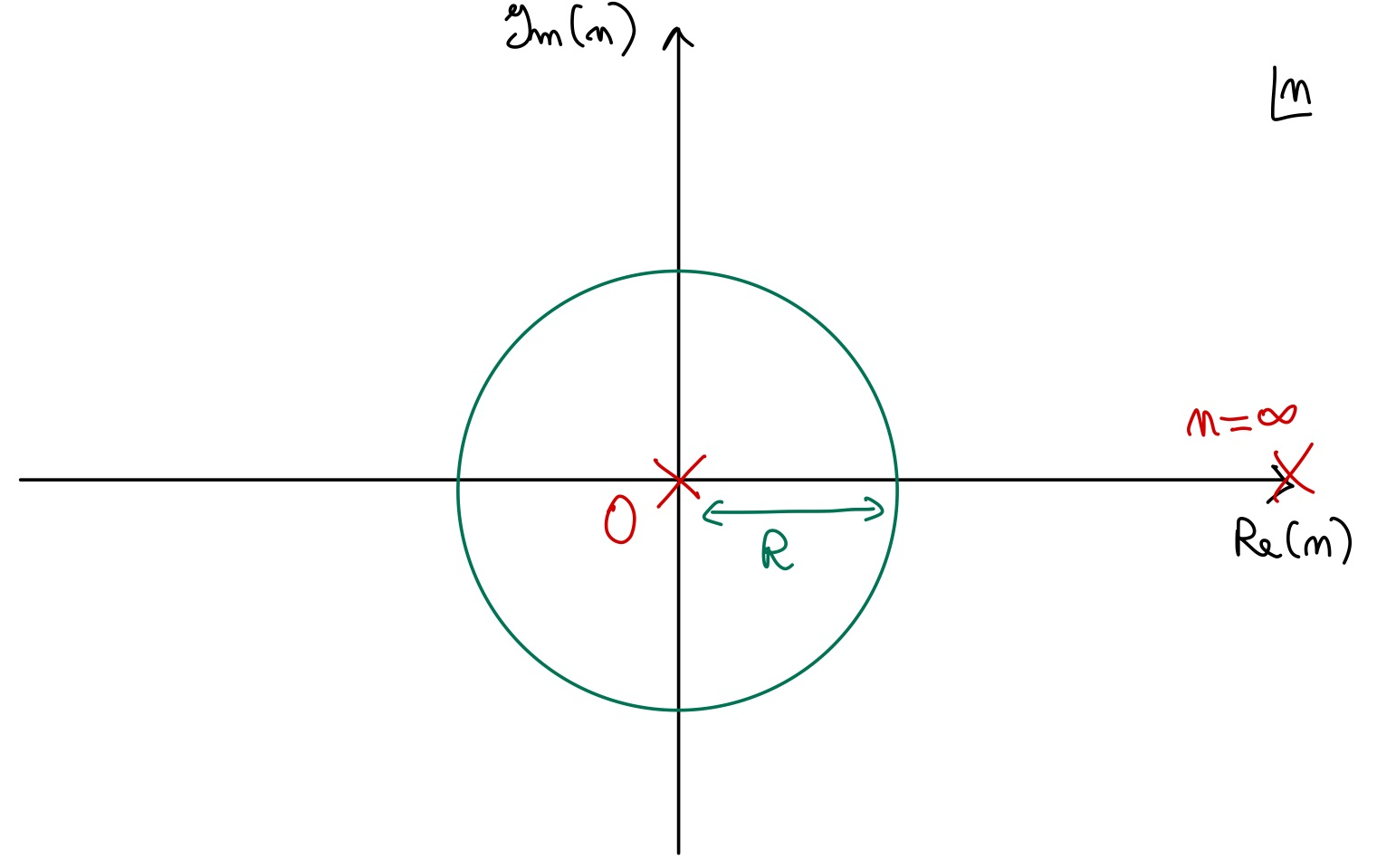}}
    \caption{(a) Closed integration contour for the evaluation of the inverse Laplace transform, composed by the vertical green line $\mathcal{C}$, plus the violet semicircle corresponding to the $I_{\subset}$ contribution. 
    (b) Deformed close contour around the origin with radius $R$. }
\end{figure}

In dimensions $d=2,3$, we can evaluate the inverse Laplace transform for the SUSY case exactly; no saddle point approximation is required. 
To that end, we observe that the integrand entering eq.~\eqref{eq:DOS_tilde} takes the following form in two and three dimensions
\be\label{SusyIntegrand}
\frac{e^{f(n)}}{2 \pi i} = \frac{e^{n (E - E_0) + \eta + \frac{\kappa}{n}}}{2 \pi i},
\ee
where $\eta$ and $\kappa$ are constants.
Using the explicit expressions \eqref{eq:SUSY_Renyi_entropy} and \eqref{eq:def_E0_SUSY}, we identify
\be
E_0 = \begin{cases}
  \frac{1}{2} \sri  & \text{if $d=2$} \\
   \frac{5}{8} \sri     & \text{if $d=3$}
\end{cases}  \qquad
\eta = \begin{cases}
   0 & \text{if $d=2$} \\
   \frac{1}{4} \sri     & \text{if $d=3$}
\end{cases}  \qquad
\kappa = \begin{cases}
  \frac{1}{2} \sri  & \text{if $d=2$} \\
   \frac{3}{8} \sri     & \text{if $d=3$}
\end{cases}
\label{eq:SUSY_constants_2d3d}
\ee
When $E>E_0$, the integrand \eqref{SusyIntegrand} presents essential singularities at $n=0$ and $n=\infty$. 
We can then complement the integration contour $\mathcal{C}$ in eq.~\eqref{eq:DOS_tilde} by a semicircle at infinity in the negative half of the complex plane, as depicted in fig.~\ref{fig:plotintsusy}.
That is, we use the identity \eqref{eq:split_integrals_Isub} 
where $D(E)$ corresponds to the contribution \eqref{eq:DOS_tilde} coming from the vertical green line $\mathcal{C}$, $I_{\subset}$ has the same integrand but evaluated on the semicircle, and $I$ is the full integral over the closed contour.
The main difference, compared with fig.~\ref{subfig:int2_delta}, is that in the SUSY case there are no branch cuts along the imaginary axis, therefore we do not need to deform the shape of the semicircle. In other words, $I_{||}=0$ in eq.~\eqref{eq:split_integrals_Isub}.

After the coordinate change $n = \tilde{R} e^{i\theta}$, we find
\be
I_{\subset}= \lim_{\tilde{R} \rightarrow \infty} \int^{3\pi/2}_{\pi/2}\frac{d\theta}{2\pi} \, \tilde{R} e^\eta e^{\tilde{R} e^{i\theta} (E-E_0)+\frac{\kappa}{\tilde{R}} e^{- i\theta}+i \theta} \, .
\ee
Since in this case $\tilde{R} \to \infty$ and $\mathrm{Re} \, (e^{i\theta}) <0$, then the integral vanishes when $E > E_0$.
For arbitrary energy $E$, after following the same argument leading to eq.~\eqref{deltaFunc}, we find that $I_\subset = -\delta(E-E_0)$. 
This shows that $I + \delta(E-E_0) = D(E)$. 
Next, let us focus on evaluating $I$ using the closed contour mentioned above.

Since the integrand is an analytic function inside all the closed contour in fig.~\ref{fig:plotintsusy} (except for the essential singularity at the origin), we can use Cauchy's theorem to deform the integration region to a circle of finite radius $R$ around the singularity at $n=0$, as shown in fig.~\ref{fig:deformed_contour}. 
Employing the parametrization $n=R e^{i \theta}$, we shall conveniently pick $R$ such that 
\be
R (E-E_0) = \frac{\kappa}{R} \quad 
\Rightarrow \quad R = \sqrt{\frac{\kappa}{E-E_0}} \, .
\ee
This gives
\be
\begin{aligned}
D(E) 
&= \sqrt{\frac{\kappa}{E-E_0}}\frac{e^\eta }{2\pi}\oint d\theta e^{2\cos(\theta)\sqrt{\kappa(E-E_0)}+i \theta} +\delta(E-E_0)
\\
& = \sqrt{\frac{\kappa}{E-E_0}}e^\eta I_1(2 \sqrt{\kappa(E-E_0)}) +\delta(E-E_0) \, .
\end{aligned}
\label{eq:SUSY_int_2d3d}
\ee
In $d=2$, using the specific constants in eq.~\eqref{eq:SUSY_constants_2d3d}, we reproduce the Calabrese-Lefevre formula \eqref{CLformulaBessel}. This is to be expected since the supersymmetric R\'enyi entropy \eqref{eq:holo_Renyi_entropy_SUSY} reduces to the conventional R\'enyi entropy \eqref{eq:holo_Renyi_entropy} when $d=2$. 
In $d=3$, we emphasize however that we obtain a novel exact result.

Even though the behavior with respect to the energy in 2d and 3d is the same, in 3d there is an additional exponential dependence on the volume (via the AdS Rindler entropy \eqref{Srindler} entering the expression of $\eta$). 
Finally, we stress that the technical reason why this analytic approach can be performed for a SUSY-invariant CFT, compared to the uncharged case in section~\ref{ssec:DOS_from_Renyi}, is that $x_n$ in eq.~\eqref{eq:xn_SUSY} and the R\'{e}nyi entropy \eqref{eq:holo_Renyi_entropy_SUSY} take a simple form.
This allowed us to directly perform the integration \eqref{eq:SUSY_int_2d3d}, that otherwise would have included a branch cut.

\subsection{Analytic results at large energies in general dimension}
\label{ssec:SUSY_largeE_anyd}

Next, we analytically study the DOS in general dimension.
Starting from the definition
\be
f(n) := E n + (1-n) S_n 
=n\left[E-\frac{\sri}{2}\left(1+2x_n-x_n^2-2x_n^{d-1}\right)\right] \, ,
\label{eq:fn_def}
\ee
and using the property 
\be
\partial_n x_n = \frac{1}{n^2 (1-d)} \, ,
\ee
we obtain 
\begin{equation}
    f'(n)
    =
    E - \frac{\sri}{2}\left[
    1+2x_\infty - 2x_\infty x_n + x_n^2 - 2(d-2)x_n^{d-2} + 2(d-2)x_n^{d-1}
    \right] \, ,
    \label{eq:fprime_SUSY_onlyX}
\end{equation}
where \eqref{eq:xn_SUSY} has been used to replace every instance of $n$ by $x_n$.

The saddle points are defined by the condition $f'(n_*)=0$, which results in
\begin{align}
    \underbrace{\frac{2E}{\sri}-1-2x_\infty}_{B}
    &=
    2(d-2)(x_{n_*}^{d-1}-x_{n_*}^{d-2}) + x_{n_*}^{2} - 2x_\infty x_{n_*} \, ,
    \label{eq:saddleX_susy}
\end{align}
where we conveniently renamed the quantity on the left-hand side as $B$.
When $d=2$, this relation reduces to the uncharged case \eqref{saddlesX}, just as the SUSY R\'{e}nyi entropy \eqref{eq:holo_Renyi_entropy_SUSY_sec2} reduces to \eqref{RenyiAndXn}.
Hence, we conclude that the two-dimensional configuration with three-dimensional bulk uncharged black hole is automatically supersymmetric-invariant.
Since we already evaluated this case in section~\ref{ssec:DOS_2d}, we now restrict the present analysis to higher dimensions $d>2$.

For $d\neq 2$, we can conveniently solve eq.~\eqref{eq:saddleX_susy} for the highest power of $x_{n_*}$
\begin{align}
    x_{n_*}^{d-1}
   &=
   \frac{1}{2(d-2)}\left[B+2x_\infty x_{n_*}-x_{n_*}^{2}+2(d-2)x_{n_*}^{d-2}\right] \, .
\end{align}
Plugging this expression into $f$ yields
\begin{align}
     f(n_*)
    &=
    n_*\frac{\sri}{2}\left[
    \frac{B}{x_\infty}+2x_\infty-2x_{n_*}x_\infty+\frac{d-3}{d-2}x_{n_*}^2+2x_{n_*}^{d-2} 
    \right] \, .
    \label{eq:fsaddle_susy}
\end{align}
For convenience, we re-write the relevant quantity $B$ (which determines the saddle points via \eqref{eq:saddleX_susy}) as follows:
\begin{equation}
    B
    =
  \cE_{\rm s}(d) \, \frac{E-E_0}{E_0}-\frac{r_{\rm cr}^2}{L^2} + (d-2) \frac{m_{\rm cr}}{L^{d-2}} \, ,
  \label{eq:expansion_B}
\end{equation}
where we used the definitions of critical mass and radius introduced for a BPS black hole in eq.~\eqref{eq:critical_mass_susy}.
This identity is the supersymmetric counterpart of eq.~\eqref{energyDensityA}, since it expresses $B$ in terms of the relative difference of energies, plus a residual contribution related to the critical parameters of the black hole.
In order to perform a large-energy expansion in the regime \eqref{eq:assumptions_cutoff} we then require $B$ to be large, and thus eq.~\eqref{eq:saddleX_susy} can only be solved if $x_{n_*} \gg 1$.
This approach leads to two possible dominant terms in the right-hand side of eq.~\eqref{eq:fsaddle_susy}, depending on the dimension $d.$

\paragraph{Large energies for $d=3.$} 
Given the exact expression \eqref{eq:SUSY_int_2d3d}, one need only expand for large energies to arrive at
\begin{equation}
   D(E)
   =
   \sqrt{\frac{5}{12 \pi}} \le  \frac{E_0}{(E-E_0)^3}  \ri^{\frac{1}{4}}   
    \exp \le \sqrt{\frac{12}{5} E_0 (E-E_0)} \ri
    \label{eq:DSUSY3d}
\end{equation}
We have checked that our approach described above to derive the density of states yields the same result. 

\paragraph{Large energies for $d>3.$}
In higher dimensions $d>3$, the dominant term in the right-hand side of eq.~\eqref{eq:saddleX_susy} is $ x_{n_*}^{d-1}$, therefore we can simplify the saddle point condition to
\begin{equation}
    B=2(d-2)x_{n_*}^{d-1}
    \qquad
    \Rightarrow
    \qquad
    x_{n_*}^{(k)}
    =
    \exp[2\pi i\frac{k}{d-1}]\left(\frac{B}{2(d-2)}\right)^{1/(d-1)}\label{eq:xd4_susy}
\end{equation}
where $k=0,1,\dots,d-2$. 
These are $d-1$ roots, while in the uncharged case \eqref{eq:xnstar_solution_largeE} we had $d$ roots.
The explicit solution for $n_*$ reads
\be
\begin{aligned}
    n_*^{(k)}
    =
    \frac{1}{(d-1)x_{n_*}-(d-2)}
    &\approx
    \frac{1}{(d-1)x_{n_*}} \\
    &=
    \frac{1}{d-1}\exp[-2\pi i\frac{k}{d-1}]\left(\frac{2(d-2)}{B}\right)^{1/(d-1)} \, .
\end{aligned}
\ee
Since $x^{d-1}_n \propto B$, all smaller powers are sub-leading, so that only the first term in eq.~\eqref{eq:fsaddle_susy} needs to be retained.\footnote{This step is different in the case $d=3$, because there is another dominant term in $x_n$, since the $\frac{d-3}{d-2}x_{n_*}^2$ term drops out in \eqref{eq:fsaddle_susy}. Indeed, the numerical factors in the argument of the square root in the exponential of eq.~\eqref{eq:DSUSY3d} 
cannot be simply obtained by setting $d=3$ in eq.~\eqref{eq:fSUSY}.}
We find
\be
\begin{aligned}
    f(n_*^{(k)})
    \approx
    n_*^{(k)} \frac{\sri}{2}\frac{B}{x_\infty}
    &=
    \sri\exp[-2\pi i\frac{k}{d-1}]\left(\frac{B}{2(d-2)}\right)^{(d-2)/(d-1)} \\
    & \approx  \exp[-2\pi i\frac{k}{d-1}]\left[ \frac{2E_0}{\mathcal{E}_{\rm s}(d)} \le \frac{E-E_0}{d-2} \ri^{d-2} \right]^{\frac{1}{d-1}} \, ,
    \label{eq:fSUSY}
\end{aligned}
\ee
where the definition of $x_\infty$ in eq.~\eqref{eq:range_xn_susy} and the approximation $B\approx  \cE_{\rm s}(d) \, \frac{E-E_0}{E_0}$ of eq.~\eqref{eq:expansion_B} were employed. 

Following similar arguments as outlined in appendix~\ref{ssec:dominant_saddles}, we find that the dominant contribution at high energies is given by the real and positive solution ($k=0$), which does not share its real part with any other saddle.
In this case, we also employ the same saddle point $n_*^{(0)}$ to compute the second derivative $f''$, which reads
\be
f''(n_*^{(0)}) \approx 
(d-1)^2 \left[ \frac{\mathcal{E}_{\rm s}(d)}{2E_0 (d-2)} (E-E_0)^d \right]^{\frac{1}{d-1}} > 0 \, .
\ee
This guarantees that the saddle-point approximation is well-defined.

\paragraph{Density of states.}
In summary, we obtained the following leading behaviour at high energies for the density of states:
\be
D(E) \approx \begin{cases}
 \le \frac{E_0}{(4 \pi)^2 (E-E_0)^{3}}\ri^{\frac{1}{4}}
\, \exp \le  2 \sqrt{ E_0 (E-E_0)}  \ri  & \text{if $d=2$} \\
    \sqrt{\frac{5}{12 \pi}} \le  \frac{E_0}{(E-E_0)^3}  \ri^{\frac{1}{4}}   
    \exp \le \sqrt{\frac{12}{5} E_0 (E-E_0)}  \ri &  \text{if $d=3$} \\
  \frac{1}{\sqrt{2\pi} (d-1)} \left[ \frac{\mathcal{E}_{\rm s}(d)}{2E_0 (d-2)} (E-E_0)^d \right]^{-\frac{1}{2(d-1)}} \,   \exp \left[ \left( \frac{2E_0}{\mathcal{E}_{\rm s}(d)}  \right)^{\frac{1}{d-1}} \le \frac{E-E_0}{d-2} \ri^{\frac{d-2}{d-1}}   \right]  & \text{if $d>3$}
\end{cases}
\label{eq:leading_DOS_susy}
\ee
Expressing these quantities in terms of the micro-canonical entropy, we get the following important result, valid in the SUSY case:
\be
\boxed{
S \propto \log D(E) =
\begin{cases}
  2 \sqrt{ E_0 (E-E_0)} + \dots  & \text{if $d=2$} \\
 \sqrt{\frac{12}{5} E_0 (E-E_0)}  + \dots &  \text{if $d=3$} \\
 \left[ \frac{2E_0}{\mathcal{E}_{\rm s}(d)} \le \frac{E-E_0}{d-2} \ri^{d-2} \right]^{\frac{1}{d-1}} + \dots  & \text{if $d>3$}
\end{cases} }
\label{eq:susy_S}
\ee
Remarkably, this result provides a generalization of the Cardy formula to supersymmetric-invariant theories. We remind the reader that this is however the density of state integrated over the charge, as per \eqref{eq:Dgamma}.
The scaling between entropy and energy coincides with the uncharged case \eqref{CardyDeformation} only when $d=2$, while it differs in other dimensions.
To our knowledge our result in $d=3$ is new, even when discussing regular CFT spectra. In general, we note that the microcanonical entropies \eqref{eq:susy_S} are smaller than those of the non-supersymmetric case \eqref{CardyDeformation}.

\subsection{Numerical results}
\label{ssec:SUSY_numerical}

\begin{figure}[ht]
    \centering
    \subfigure[]{\includegraphics[scale=0.69]{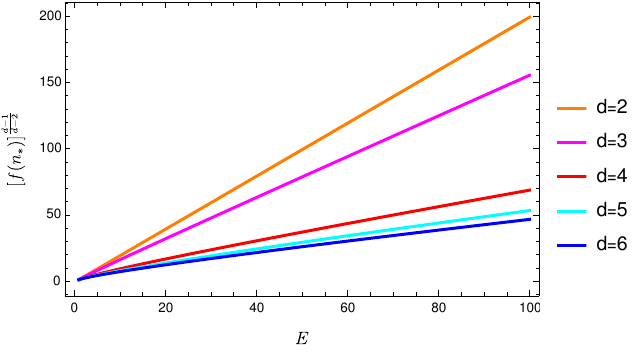}}
    \subfigure[]{\includegraphics[scale=0.69]{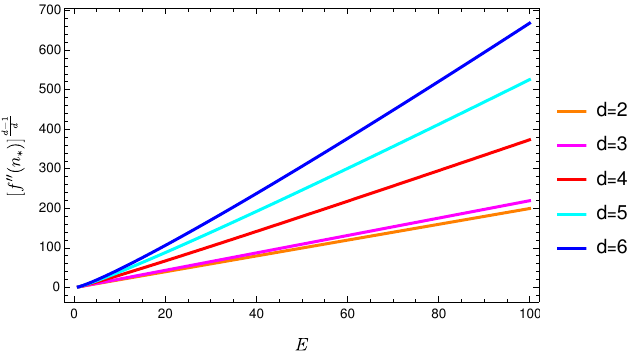}}
    \caption{Plot of $f^{\frac{d-1}{d-2}}$ and $(f'')^{\frac{d-1}{d}}$ at the dominant saddle point as functions of the energy $E$, for fixed $\sri=1$. The parameter $d$ refers to the spacetime dimensions of the CFT. When $d=2,3$, we plot $f^2$ and $(f'')^{\frac{2}{3}}$ instead.    }
    \label{fig:plotfsusy}
\end{figure}

We supplement the analytic computation performed in section~\ref{ssec:SUSY_largeE_anyd} with a numerical evaluation of the DOS.
Other than providing a consistency check of the previous results, it presents the advantage of not relying on the large energy limit \eqref{eq:assumptions_cutoff}, but only on the saddle point approximation.
We follow the same method outlined in section~\ref{ssec:numerical_analysis}, with the difference that the leading scaling of the DOS with $E$ at high energies reads
\be
f(n_*) \sim \begin{cases}
    \sqrt{E} & \text{if $d=2,3$}  \\
    E^{\frac{d-2}{d-1}} & \text{otherwise} 
\end{cases}  \qquad
f''(n_*) \sim \begin{cases}
    E^{\frac{2}{3}} & \text{if $d=2,3$}  \\
    E^{\frac{d}{d-1}} & \text{otherwise} 
\end{cases} 
\ee
We report in fig.~\ref{fig:plotfsusy} a numerical plot of $f^{\frac{d-1}{d-2}}$ and $(f'')^{\frac{d-1}{d}}$ ($f^2$ and $(f'')^{\frac{2}{3}}$ in dimensions $d=2,3$) to show that they approach a linear growth for large energies, thereby confirming our analytic results.

%% file: Draft_Sections/ShapeDeformations.tex
\section{Shape deformations}
\label{sec:shape_def_DOS}

The goal of this section is to extend the analysis of the density of states performed in sections~\ref{sec:holographic_CFT_uncharged} and \ref{subsec:DOS_SUSY} for holographic CFTs to the case where a small shape deformation of a spherical (or planar) entangling surface is performed.
We begin with some general observations, and then focus on the setting with vanishing chemical potential in section~\ref{ssec:shapedef_uncharged}, and with supersymmetric invariance in section~\ref{ssec:shapedef_susy}.

Consider the reduced density matrix $\rho_A$ describing the entanglement between a subregion $A$ and its complement $\bar{A}$ on an arbitrary time slice of a CFT.
We consider the case when the boundary $\partial A$ separating the two subregions is composed of a codimension-two entangling surface $\Sigma$ whose profile is infinitesimally close to a spherical (or planar) shape, as defined by eq.~\eqref{deformation}.
The identity \eqref{eq:variation_Sn_in_terms_of_CD} implies that the leading-order variation of the R\'{e}nyi entropy reads
\be
(1-n) \delta S_n = \mathfrak{b} C_D (n)  + \mathcal{O} (f^4)  \, , \qquad
\mathfrak{b} := 
 \frac{1}{2}  
 \int_{\Sigma} dw \int_{\Sigma} dw' \, 
\frac{f^a (w) f_a (w')}{(w-w')^{2(d-1)}} \, ,
\label{eq:definition_b}
\ee
where we introduced the quantity $\mathfrak{b}$ for convenience, and we used the fact that $C_D(n=1)=0$ for both the uncharged and SUSY-invariant cases.
In the previous formula, it is important to observe that:
\begin{itemize}
    \item The series expansion starts at order $\mathfrak{b} \sim \mathcal{O}(f^2)$ with $f^a \sim \mathcal{O}(\varepsilon)$ being the deformation profile, since the one-point function of the displacement operator vanishes.
    Here, $\varepsilon$ represents a small parameter.
    \item The dependence on the replica index $n$ is completely encoded in the coefficient $C_D$.
    \item The dependence on the precise shape is contained inside $\mathfrak{b}$. 
    This quantity is UV-divergent, but its universal (\ie scheme-independent) contributions amount to a finite part in odd dimensions $d$, and to a logarithmic term in the cutoff in even dimensions \cite{Bianchi:2015liz}.\footnote{We assume that $\mathfrak{b}$ is renormalized using a regulator that preserves conformal invariance. This gets rid of power-law UV divergences.}
\end{itemize}
We aim to compute the consequences of this deformation on the DOS.
Following steps similar to section~\ref{ssec:DOS_from_Renyi}, we define the DOS as the inverse Laplace transform
\be
D_{b}(E) = \frac{1}{2 \pi i} \int_{\mathcal{C}} 
dn \, e^{nE} e^{(1-n) (S_n +\delta S_n)}  \, ,
\label{eq:deformed_DOSb}
\ee
where the contour $\mathcal{C}$ now runs on the right of all the singularities of $e^{(1-n) (S_n +\delta S_n)}$, and the subscript in $D_{\mathfrak{b}}(E)$ denotes the presence of a deformation with profile controlled by the quantity $\mathfrak{b}$ in eq.~\eqref{eq:definition_b}.
In order to extract the density of states, we need to study the exponent
\be
f(n) := E n + (1-n) S_n + (1-n) \delta S_n \, ,
\label{eq:deff_deformations}
\ee
defined from the identity
\begin{equation}
\label{eq:DOScontour_shapedef}
    D_{\mathfrak{b}}(E) 
    =
    \lim_{K\to\infty}\int_{r - i K}^{r + i K} \frac{dn}{2\pi i} \,  e^{n E} e^{(1-n)(S_n+ \delta S_n)}
    :=
    \lim_{K\to\infty}\int_{r - i K}^{r + i K} \frac{dn}{2\pi i} \, e^{f(n)}     \, .
\end{equation}
It is important to notice that the analytic results available in the case of shape deformations (reviewed in section~\ref{ssec:shape_def_review}) rely on the assumption that $\mathfrak{b}$ is small, in the precise sense described by the inequality \eqref{eq:assumptions_cutoff2} below.
In this regime, the analytic structure of the integrand in eq.~\eqref{eq:DOScontour_shapedef} is still determined by the discussion outlined around eq.~\eqref{eq:branch_points}, therefore the inverse Laplace transform is well-defined if we take $r>0$.  

In the following, we determine $D_{\mathfrak{b}}(E)$ with a saddle point approximation for the uncharged and the SUSY-invariant cases.
In the latter setting, it is implicitly understood that the DOS is defined with the same procedure as outlined in section~\ref{ssec:which_DOS_SUSY}.

\subsection{Uncharged case}
\label{ssec:shapedef_uncharged}

Let us begin with the uncharged case.
The relevant exponent \eqref{eq:DOScontour_shapedef} becomes
\be
f(n)  = n \left[ E - \sri \le 1 - \frac{1}{2} (1+x_n^2) x_n^{d-2} \ri \right]
+ \mathfrak{b} C_D (n) \, .
\label{eq:fn_shapedef1}
\ee
By using the identities
\be
n= \frac{2 x_n}{d x_n^2 - (d-2)} \, , \qquad
\p_n x_n = \frac{x_n}{n (1-d n \, x_n)} = - \frac{\left[ 2 + d (x_n^2 -1) \right]^2}{2 \le d x_n^2 + d -2 \ri} \, ,
\label{eq:identities_n_xn}
\ee
we find the derivative
\be
f'(n) = E - \frac{2E_0}{\mathcal{E}(d)} \left[1+\frac{(d-1)}{2}(x_n^d-x_n^{d-2})\right]  + \mathfrak{b} \, \frac{\partial C_D}{\partial n} \, ,
\label{eq:derfn_shapedef1}
\ee
where we used the latter identity in eq.~\eqref{Srindler} to express the thermal entropy of Rindler AdS in terms of the minimal eigenvalue of the modular Hamiltonian in the case of spherical (planar) entangling surface.
In other words, here $E_0$ denotes the lowest eigenenergy of the undeformed system.

In order to make progress analytically, we consider the regime
\begin{equation}
   \frac{\mathfrak{b} C_T}{E_0} \ll 
    1  \ll  \frac{E-E_0}{E_0} \, ,
    \label{eq:assumptions_cutoff2}
\end{equation}
which corresponds to a limit of large energies and small deformations.
An analysis similar to section~\ref{ssec:computation_DOS} shows that a solution for the saddle points $n_*$ such that $f'(n_*)=0$ can only be achieved when $x_n \gg 1$, or equivalently when $n \ll 1$.
In the limit $n \rightarrow 0$, an analytic expansion of the displacement norm $C_D$ was found in reference \cite{Bianchi:2016xvf}, as we reported in eq.~\eqref{eq:small_CD}.

Truncating the Laurent series for $C_D$ around $n=0$ at $\mathcal{O}(n)$, we get 
\be
\frac{\partial C_D}{\partial n} \approx 2^{d-1} \pi^2 C_T \, \frac{d-1}{d^{d-1} (d+1)}  \frac{1}{n^d} \approx 
\frac{\pi^2 C_T d (d-1)}{2 (d+1)} \le  \frac{2+d(x_n^2-1)}{d x_n}  \ri^d 
\, ,
\ee
where we used eq.~\eqref{eq:identities_n_xn} in the last step.
Plugging this expression inside eq.~\eqref{eq:derfn_shapedef1} and solving at leading order for the saddle points, we obtain
\be
x_{n_*}^{(k)} 
= e^{2\pi i k/d}  \left[  \frac{m(E)}{L^{d-2}} \frac{1}{ 1 - \frac{\mathfrak{b} \mathcal{E}(d)}{2E_0}  \frac{\pi^2 d C_T}{d+1}}  \right]^{\frac{1}{d}}
\approx e^{2\pi i k/d} \left[ \frac{\mathcal{E}(d)}{(d-1) \le  1 - \frac{\mathfrak{b} \mathcal{E}(d)}{2E_0}  \frac{\pi^2 d C_T}{d+1} \ri}  \frac{E-E_0}{E_0 }  \right]^{\frac{1}{d}}   \, ,
\label{eq:solution_shapedep_smalln}
\ee
where we used the identity \eqref{energyDensityA} in the first step, and the assumption \eqref{eq:assumptions_cutoff2} to achieve the final approximation.
Comparing with eq.~\eqref{eq:xnstar_solution_largeE}, we notice that the shape deformation amounts to a shift of the 
denominator. 
In the previous result, only the leading order in an expansion for small $\mathfrak{b}$ is trustworthy.
From now on, we focus on the only real and positive root, in other words we take $k=0$ in eq.~\eqref{eq:solution_shapedep_smalln}.

We approximate the exponent $f(n)$ and its second derivative for large $x_n$ as follows:
\begin{subequations}
\be
f(n) \approx \frac{2}{d x_n} \left[ E + \frac{E_0}{\mathcal{E}(d)}  \le 1 - \frac{\mathfrak{b} \cE(d)}{2E_0} \frac{\pi^2 d C_T}{d+1}  \ri x_n^d \right] \, ,
\ee
\be
f''(n) \approx  \frac{d^2(d-1)}{2} \frac{E_0}{\mathcal{E}(d)} \le 1 - \frac{\mathfrak{b} \cE(d)}{2E_0} \frac{\pi^2 d C_T}{d+1}  \ri  x_n^{d+1}   \, .
\ee
\end{subequations}
In the regime \eqref{eq:assumptions_cutoff2}, we conclude that $f''(n) \geq 0$ when $n \geq 0$. 
This confirms that the saddle point that we obtained is well-behaved.
When evaluating $f(n)$ and $f''(n)$ on the real and positive solution to eq.~\eqref{eq:solution_shapedep_smalln}, we obtain
\begin{subequations}
\be
f(n_*) 
 \approx 2 \left[  \frac{E_0}{\cE(d)} \le 1 - \frac{\mathfrak{b} \cE(d)}{2E_0} \frac{\pi^2 d C_T}{d+1}  \ri  \le \frac{E-E_0}{d-1} \ri^{d-1}  \right]^{\frac{1}{d}}  \, ,
\ee
\be
f''(n_*) 
 \approx  
  \frac{d^2}{2} \left[  \frac{\cE(d)}{(d-1) E_0} \le 1 - \frac{\mathfrak{b} \cE(d)}{2E_0} \frac{\pi^2 d C_T}{d+1}  \ri^{-1}  \le E-E_0 \ri^{d+1}  \right]^{\frac{1}{d}}  \, ,
\ee
\end{subequations}
valid in the regime \eqref{eq:assumptions_cutoff2}.
Consequently, the DOS \eqref{eq:deformed_DOSb} in the presence of small shape deformations of the entangling surface away from a spherical (planar) shape reads
\be
\begin{aligned}
D_{\mathfrak{b}} (E) & \approx \frac{1}{d \sqrt{\pi}} 
\left(\frac{\cE(d)(E-E_0)^{d+1}}{E_0(d-1)}\right)^{-\frac{1}{2d}} 
\le 1 - \frac{\mathfrak{b} \cE(d)}{2E_0} \frac{\pi^2 d C_T}{d+1}  \ri^{\frac{1}{2d}}  \\
& \times  \exp \left[ 2 \le \frac{E_0}{\cE(d)} \ri^{\frac{1}{d}} \le 1 - \frac{\mathfrak{b} \cE(d)}{2E_0} \frac{\pi^2 d C_T}{d+1}  \ri^{\frac{1}{d}}  \le \frac{E-E_0}{d-1} \ri^{\frac{d-1}{d}}  \right]  \, . 
\end{aligned}
\label{eq:DOS_shape_def_result}
\ee
Clearly, the limit $\mathfrak{b} \rightarrow 0$ leads to the undeformed result \eqref{eq:leading_DOS_largeE}.
The micro-canonical entropy is finally given by
\be
\boxed{
S(E) \approx  2 \le \frac{E_0}{\cE(d)} \ri^{\frac{1}{d}} \le 1 - \frac{\mathfrak{b} \cE(d)}{2E_0} \frac{\pi^2 d C_T}{d+1}  \ri^{\frac{1}{d}}  \le \frac{E-E_0}{d-1} \ri^{\frac{d-1}{d}}  + \dots }
\ee
Interestingly, we notice that the existence of a shape deformation only affects the prefactors, but does not modify the scaling of the microcanonical entropy with the energy.
It would be interesting to further investigate whether this feature is universal and robust, or it only holds in the regime of small deformations.

\subsection{Supersymmetric case}
\label{ssec:shapedef_susy}

In the supersymmetric case, an analytic evaluation of $f(n)$ in eq.~\eqref{eq:deff_deformations} can be achieved by using the conjecture 
\eqref{eq:conj_CD}, which was shown to hold for holographic theories in \cite{Baiguera:2022sao}.
For later convenience, we record the following simplification
\be
\pi d \, \Gamma \le \frac{d+1}{2} \ri \le \frac{2}{\sqrt{\pi}} \ri^{d-1} \le \frac{L}{\ell_{\mathrm{P}}} \ri^{d-1} =
\pi^2 \frac{d-1}{d+1} C_T \, ,
\label{eq:definition_a}
\ee
where in the last step we used the definition \eqref{eq:def_CT}.
Plugging the latter result and the identity \eqref{eq:definition_b} inside the exponent \eqref{eq:deff_deformations}, we get
\be
f(n) = n \left[E-\frac{E_0}{\mathcal{E}_{\rm s}(d)}\left(1+2x_n-x_n^2-2x_n^{d-1}\right) + 2 \pi^2 \frac{d-1}{d+1} \, \mathfrak{b} C_T  x_n^{d-2}  \le 1 - x_n  \ri \right] \, ,
\label{eq:fn_deformations_SUSY}
\ee
where we used the definition \eqref{eq:def_E0_SUSY} of the minimal eigenenergy of the modular Hamiltonian in the undeformed case. 
Now we apply the identities 
\be
 n=\frac{1}{(d-1)x_n-(d-2)} \, , \qquad
 \p_n x_n = \frac{1}{n^2 (1-d)}  = \frac{\left[ (d-1)x_n-(d-2) \right]^2}{1-d} \, , 
\ee
to compute
\be
\begin{aligned}
    f'(n) & = E  - \frac{E_0}{\mathcal{E}_{\rm s}(d)}\left[
    1+2x_\infty - 2x_\infty x_n + x_n^2 - 2(d-2)x_n^{d-2} + 2(d-2)x_n^{d-1}
    \right] \\
    & + \frac{2 \pi^2 \frac{d-1}{d+1} \, \mathfrak{b} C_T }{d-1} \, x_n^{d-3} \left[  4 + d^2 (x_n-1)^2 -5 x_n + 2 x_n^2 -d (3x_n-4)(x_n-1)  \right] \, .
\end{aligned}
\ee
As usual, the saddle points are determined by the conditions $f'(n)=0$.
In the high-energy regime \eqref{eq:assumptions_cutoff}, this requires to look for solutions with $x_n \gg 1$.
Similar to the undeformed case considered in section~\ref{ssec:SUSY_largeE_anyd}, the dominant term inside $f'(n)$ depends on the number of spacetime dimensions $d$.

Let us analyze the various cases separately.
When $d=2$, there do not exist non-trivial deformations, since the codimension-two entangling surface is only composed by two points.

\paragraph{Case $d=3.$}
When $d=3$, the only real and positive saddle point is given by
\be
x_{n_*} = \sqrt{ \frac{5}{12} \frac{E-E_0}{E_0} \frac{1}{1- \frac{5 \pi^2 \, \mathfrak{b} C_T }{12 E_0}}} \, .
\ee
which implies
\begin{subequations}
    \be
f(n_*) \approx  \sqrt{  \frac{12}{5} E_0 \le 1- \frac{5 \pi^2 \, \mathfrak{b} C_T }{12 E_0} \ri (E-E_0) } \, ,
\ee
\be
f''(n_*) = \frac{1}{2} \le \frac{3}{2} \sri -  \pi^2 \, \mathfrak{b} C_T  \ri \le 2x_{n_*} -1 \ri^3  \approx \frac{4 (E-E_0)^{\frac{3}{2}}}{ \sqrt{\frac{12}{5} E_0 \le 1- \frac{5 \pi^2 \, \mathfrak{b} C_T }{12 E_0} \ri} } \, .
\ee
\end{subequations}

\paragraph{Case $d>3.$}
In higher dimensions, the saddle points are determined by
\be
x_{n_*} = \left[ \frac{\cE_{\rm s}(d)}{d-2} \frac{E-E_0}{2 E_0}  \le 1 - \frac{\pi^2 (d-1) \mathfrak{b} C_T \cE_{\rm s} (d)}{(d+1) E_0} \ri^{-1} \right]^{\frac{1}{d-1}} \, .
\ee
The dominant saddle point corresponds to the real and positive root.
In this way, we find
\begin{subequations}
\be
f(n_*) \approx \left[ \frac{2E_0}{\cE_{\rm s} (d)} \le 1 - \frac{\pi^2 (d-1) \mathfrak{b} C_T \cE_{\rm s} (d)}{(d+1) E_0} \ri \le \frac{E-E_0}{d-2} \ri^{d-2} \right]^{\frac{1}{d-1}}  \, ,
\ee
\be
    f''(n_*) \approx (d-1)^2  \left[ \frac{\cE_{\rm s}}{2E_0 (d-2)}  \le 1 - \frac{\pi^2 (d-1) \mathfrak{b} C_T \cE_{\rm s} (d)}{(d+1) E_0} \ri^{-1}  (E-E_0)^d \right]^{\frac{1}{d-1}} \, .
\ee
\end{subequations}

\paragraph{Summary.}
In various dimensions, the DOS in the presence of a small shape deformation of the entangling surface of a SUSY-invariant CFT reads
\be
D_{\mathfrak{b}}(E)  \approx 
\begin{cases}
  \left[  \frac{3}{80 \pi^2} E_0 \le 1- \frac{5 \pi^2 \, \mathfrak{b} C_T }{12 E_0} \ri  \right]^{\frac{1}{4}} \, (E-E_0)^{- \frac{3}{4}}  \,   & \text{if $d=3$}  \\
\times \exp \left[   \sqrt{  \frac{12}{5} E_0 \le 1- \frac{5 \pi^2 \, \mathfrak{b} C_T }{12 E_0} \ri (E-E_0) }  \right]  & \\
  \frac{1}{\sqrt{2\pi} (d-1)} \left[ \frac{\cE_{\rm s}}{2E_0 (d-2)}  \le 1 - \frac{\pi^2 (d-1) \mathfrak{b} C_T \cE_{\rm s} (d)}{(d+1) E_0}  \ri^{-1}  (E-E_0)^d \right]^{-\frac{1}{2(d-1)}}    & \text{if $d>3$} \\
  \times \exp \left[ \le \frac{2E_0}{\cE_{\rm s} (d)} \ri^{\frac{1}{d-1}} \le 1 - \frac{\pi^2 (d-1) \mathfrak{b} C_T \cE_{\rm s} (d)}{(d+1) E_0}  \ri^{\frac{1}{d-1}} \le \frac{E-E_0}{d-2} \ri^{\frac{d-2}{d-1}} \right]   & \\
\end{cases}  
\label{eq:DOS_susy_shapedef}
\ee
where we neglected the two-dimensional case since there are only trivial deformations.
The corresponding microcanonical entropy is given by
\be
\boxed{ S(E) \approx 
\begin{cases}
      \sqrt{  \frac{12}{5} E_0 \le 1- \frac{5 \pi^2 \, \mathfrak{b} C_T }{12 E_0} \ri (E-E_0) }   + \dots & \text{if $d=3$} \\
       \le \frac{2E_0}{\cE_{\rm s} (d)} \ri^{\frac{1}{d-1}} \le 1 - \frac{\pi^2 (d-1) \mathfrak{b} C_T \cE_{\rm s} (d)}{(d+1) E_0}   \ri^{\frac{1}{d-1}} \le \frac{E-E_0}{d-2} \ri^{\frac{d-2}{d-1}} + \dots & \text{if $d>3$}
\end{cases}}
\ee
Similar to the uncharged case, we notice that the effect of a small shape deformation only affects the overall prefactor, but does not change the energy scaling.

%% file: Draft_Sections/Conclusion.tex
\section{Conclusions}
\label{sec:conclusions}

\subsection{Discussion of the results}

We studied the density of states of the modular Hamiltonian for $d$-dimensional holographic CFTs.
The strategy was to import the holographic results for the R\'{e}nyi entropies previously computed in the literature (\eg see \cite{Casini:2011kv,Hung:2011nu,Belin:2013uta}), and compute the density of states as the inverse Laplace transform of the partition function.

We always worked under the assumption of large central charge $C_T \gg 1$ (equivalently, $G_N \rightarrow 0$), such that the holographic results for the R\'{e}nyi entropy could be trustfully imported.
In this regime, we determined the entanglement spectrum via a saddle point approximation of the holographic partition function.
Most of this paper focused on the high-energy regime $\frac{E-E_0}{E_0} \gg 1$, where we could analytically compute the density of states; small or intermediate energies were numerically investigated.
A summary of the analytic results for the density of states is collected in table~\ref{tab:results}.
In particular, we considered the following cases:
\begin{itemize}
    \item Reading by columns: first, holographic CFTs dual to topological black holes with vanishing charge (uncharged case).
    Second, supersymmetric-invariant holographic CFTs dual to topological charged black holes satisfying a BPS condition (SUSY case). 
    \item Reading by rows: first, configurations with a spherical (or planar) entangling surface. Second, settings where a small deformation of such shape is performed.
\end{itemize}

\begin{table}[t!]   
\begin{center}    
\begin{tabular}  {|c|c|c|} \hline  \textbf{Density of states} $D(E)$ & \textbf{Uncharged case}   & \textbf{SUSY case} \\ \hline
\rule{0pt}{4.9ex} \begin{tabular}{@{}c@{}} \textbf{Spherical (planar)} \\ \textbf{entangling surface} \end{tabular}   &  Eq.~\eqref{eq:leading_DOS_largeE}  &  Eq.~\eqref{eq:leading_DOS_susy}  \\
\rule{0pt}{4.9ex} \textbf{Shape deformations} & Eq.~\eqref{eq:DOS_shape_def_result}  &  Eq.~\eqref{eq:DOS_susy_shapedef}   \\[0.2cm]
\hline
\end{tabular}  
\caption{High-energy approximation of the density of states $D(E)$ for various physical configurations.} 
\label{tab:results}
\end{center}
\end{table}

Famously, the R\'{e}nyi entropies admit a plethora of scheme-dependent terms associated with the power-law divergences in the UV cutoff \cite{Ryu:2006ef,Nishioka:2009un}.
For holographic theories, this series expansion is captured by the volume of the dual hyperbolic space $V_{\Sigma}$.
The results collected in table~\ref{tab:results} are expressed in terms of the lowest eigenenergy of the modular Hamiltonian $E_0 = \lim_{n \rightarrow \infty} S_n$, which formally contains the divergent contribution from $V_{\Sigma}$.
In order to extract universal information from the density of states, one only needs to retain the scheme-independent part of the volume. The universal content of the results obtained in this work is also discussed from a different perspective in appendix~\ref{app:universality_Edep}. 
Alternatively, one can compute the following ratio of microcanonical entropies
\be
\mathcal{S} (E) := \frac{S(E)}{S(E_{\rm ref})} = \frac{\log D(E)}{\log D(E_{\rm ref})}  \, ,
\ee
where $E_{\rm ref} > E_0$ is a fixed reference energy.
This operation isolates the leading scaling of the microcanonical entropy with $E$ at high energies, which gives a relevant physical content.
In this way, we find 
\begin{subequations}
\begin{align}
& \mathcal{S}(E) \approx \le \frac{E-E_0}{E_{\rm ref} -E_0} \ri^{\alpha}  \, , & \\
&
\alpha|_{\rm uncharged} = \frac{d-1}{d} \, , 
\qquad
\alpha|_{\rm SUSY} = \begin{cases}
    \frac{1}{2} & \text{if $d=2,3$} \\
    \frac{d-2}{d-1} & \text{if $d>3$}
\end{cases} &
\label{eq:summary_SE_conclusions}
\end{align}
\end{subequations}
The other physical content contained inside the density of states that we computed in this paper comes from the logarithmic corrections to the micro-canonical entropy.
Their universal contribution can be extracted from the ratio
\be
\mathcal{D}(E) := \frac{D(E)}{D(E_{\rm ref})} \, ,
\ee
which reads
\begin{subequations}
\begin{align}
& \mathcal{D}(E) \approx \le \text{exp term} \ri  \le \frac{E - E_0}{E_{\rm ref} - E_0}  \ri^{\sigma} \, , & \\
&
\sigma|_{\rm uncharged} = - \frac{d+1}{2d} \, , 
\qquad
\sigma|_{\rm SUSY} = \begin{cases}
   - \frac{3}{4} & \text{if $d=2,3$} \\
   - \frac{d}{2(d-1)} & \text{if $d>3$}
\end{cases} &
\label{eq:summary_DE_conclusions}
\end{align}
\end{subequations}
The previous formulae are also valid in the presence of small shape deformations of the entangling surface.
We stress that the saddle point approximation identifies two contributions to the  microcanonical entropy.
The dominant one, reported in eq.~\eqref{eq:summary_SE_conclusions}, provides a generalization of the Cardy formula to higher dimensions.
This result agrees with the previous findings for the case of CFT spectra in two (\eg see \cite{Cardy:1986ie,Calabrese:2008iby,Hartman:2013mia,Hartman:2014oaa,Cardy:2016fqc,Tonni:2017jom,Alba:2017bgn}) and higher dimensions \cite{Verlinde:2000wg,Shaghoulian:2015lcn,Benjamin:2023qsc}.
The logarithmic corrections to the entropy, highlighted in eq.~\eqref{eq:summary_DE_conclusions}, correspond to the Gaussian fluctuations around the saddle point. 

Since the density of states reported in table~\ref{tab:results} matches with the analysis of \cite{Benjamin:2023qsc} for CFT spectra valid in the regime \eqref{eq:validity1_higherd_Cardy}, we can infer that our results apply beyond the holographic regime \eqref{eq:validity2_higherd_Cardy}, where we derived them.
However, there are two caveats. First, reference \cite{Benjamin:2023qsc} studied the case of compact CFTs, while our analysis applied to the non-compact hyperbolic space.
Second, we considered the case of entanglement spectra (rather than CFT spectra), which contains additional UV divergences coming from the edges of the subregion. We will come back to this issue in section~\ref{sec:Future}.

This paper also contains novel results in the regime $E>E_0$, without assuming large energies.
We found an analytic expression for the density of states in the supersymmetric case when $d=2,3$, as reported in eq.~\eqref{eq:SUSY_result_intro2}.
In this case, it is interesting to observe that three-dimensional SUSY theories admit a density of states proportional to a Bessel function, that in the high-energy limit reduces to the usual Cardy formula \eqref{eq:Cardy_intro}.
Finally, we numerically computed the density of states in dimensions $2 \leq d \leq 6$ at generic $E$.
In the high-energy limit, the corresponding plots fit well with the asymptotic behaviour reported in eq.~\eqref{eq:summary_SE_conclusions}.

\subsection{Future developments}\label{sec:Future}

There are several interesting problems which remain open after the study initiated in this work.
Below we summarize the main directions that we plan to study in the future:
\begin{enumerate}
    \item \textbf{Dominant saddle points.}
    In the high-energy limit, we found several saddle points over which we should sum, uniformly distributed over a circle centered at the origin of the complex plane. Throughout this work, we only retained the dominant contribution in the summation, coming from the saddle point which maximizes the integral. 
    While in lower dimensions the dominant saddle is the only positive and real solution, starting from $d>4$ there exist pairs of roots which have the same positive real part, and whose combined sum can overcome the previous contribution to the integral.
    While we stress that the contributions from multiple saddle points only affect the prefactor in the micro-canonical entropy but do not modify the universal scaling \eqref{eq:summary_SE_conclusions}, it would be interesting to understand under what conditions other saddles can dominate. This would lead to phase transitions where different dual black hole configurations are selected. 

    \item \textbf{Small energies.} It is interesting to explore the entanglement density at small energies. In this case one has to be weary of phase transitions, which might alter the form of the R\'enyi entropies. To address this point, it is reasonable to consider concrete holographic scenarios with phase transitions in order to obtain a feeling for the effect of a phase transition on the entanglement spectrum.\footnote{We thank A. Belin for discussion on this point.} 

    \item \textbf{Coarse-graining of the entanglement spectrum.}
    The entanglement spectrum \eqref{eq:DE_delta_conclusions} is extracted from a gravitational computation giving a smooth function of the modular energy $E$ for $E>E_0$.
    However, it is known that such relation is altered in the full theory, since the exact density of states is a summation of Dirac $\delta$-distributions. 
    The physical intuition is that holography automatically selects an analytic continuation of the R\'{e}nyi entropy, and this leads to an inverse Laplace transform  
    performing a coarse-graining of the full density of states.
    It would be interesting to understand the precise mechanism that establishes the relation between the coarse-grained and the exact entanglement spectra.
    The possible analytic continuation of the R\'{e}nyi entropy to the complex plane (subject to appropriate boundary conditions) are investigated by Tauberian theory, see, \eg \cite{Pappadopulo:2012jk,Qiao:2017xif,Mukhametzhanov:2019pzy}.
    
    \item \textbf{Cosmic brane interpretation.}
    Using the holographic interpretation of the R\'{e}nyi entropy proposed in \cite{Dong:2016fnf}, we found in eq.~\eqref{eq:DOS_brane} a direct relation between the density of states and the area of a dual codimension-two backreacting cosmic brane.
    Firstly, it would be relevant to test this conjecture in a simple setting, for instance in the three-dimensional SUSY case where we have at our disposal exact results similar to the Cardy formula.
    Secondly, we observe that the tension $T_n = \frac{n-1}{4 n G_N}$ has the same scaling in the replica index as the chemical potential required to achieve supersymmetry invariance. It would be interesting to understand whether this analogy has a profound meaning.
    Finally, the discussion reported in bullet 1 suggests that there could be regimes where the dominant saddle points are complex solutions sharing the same real part. In that case, the dual brane would have imaginary tension. It would be intriguing to shed light on these bulk geometric configurations. 
    \item \textbf{General chemical potential.}
    In this work, we focused for simplicity on two special choices of chemical potential: either vanishing, or fine-tuned to obtain a supersymmetric-invariant theory.
    We plan to extend our analysis to arbitrary chemical potential, whose corresponding R\'{e}nyi entropy was computed in \cite{Belin:2013uta} for a spherical (planar) entangling surface, and in \cite{Baiguera:2022sao} in the presence of small shape deformations.
    This procedure will lead to a density of states depending on both the energy of the modular Hamiltonian and the charge conjugate to the chemical potential, as in eq.~\eqref{eq:partition_function_nmu}.
    We aim to understand the meaning of this density of states over different charge sectors, and to compute it via a similar saddle point approximation.
    \item \textbf{Symmetry-resolved density of states.}
    Symmetry-resolved R\'{e}nyi entropies have received much attention recently as a tool to study features of the entanglement spectrum in the presence of charge in the context of condensed matter systems, quantum simulations and theoretical models, \eg see \cite{Goldstein:2017bua,Bonsignori:2019naz,Murciano:2019wdl,Capizzi:2020jed,Murciano:2020lqq,Azses:2020wfx,Capizzi:2021zga,Calabrese:2021wvi,Weisenberger:2021eby,Murciano:2022lsw,Azses:2022nfl,DiGiulio:2022jjd,Northe:2023khz,Murciano:2023zvk,DiGiulio:2023nvz,Kusuki:2023bsp,Castro-Alvaredo:2024azg,Bianchi:2024aim}. 
    This setting is closely related to our analysis, since the corresponding partition function is obtained via a Fourier transform of the expression \eqref{eq:general_replica_trick}.
    Since this step exchanges the fixed parameter among conjugate pairs, we expect that the corresponding density of states would be a function of the chemical potential.
    It would be interesting to learn novel properties of the entanglement spectrum as a function of this quantity, instead of the charge.
    \item \textbf{Boundary effects in entanglement spectrum.} It was pointed out in \cite{ohmori2015physics} that a proper study of entanglement requires the introduction of boundaries at the entangling edge. The authors argue that this procedure is not just a regularization scheme but provides concrete access to the Hilbert space of the subregion. This connection has been exploited successfully in a number of studies on the entanglement spectrum in two-dimensional CFT \cite{DiGiulio:2022jjd, Northe:2023khz, Kusuki:2023bsp}. Regarding the density of states of the entanglement spectrum, this approach has led to the discovery of global degeneracies in the entanglement spectrum in two dimensions \cite{Alba:2017bgn}, which are induced by Affleck-Ludwig boundary entropies. It is interesting to develop this approach also in dimensions larger than two and confirm whether additional degeneracies occur here as well and how they are characterized. Importantly, the main differences between regular CFT spectra and entanglement spectra reside in these boundary contributions, i.e., they quantify how much the entanglement spectra differ from \eqref{eq:Ooguri_intro} as obtained already for conventional CFT spectra in \cite{Benjamin:2023qsc}.
    \item \textbf{Higher-curvature gravity.} 
   In this work, we focused on holographic CFTs dual to Einstein-Maxwell gravity. One can consider holographic CFTs dual to higher-curvature bulk theories, for instance Gauss-Bonnet gravity \cite{Lovelock:1970zsf,Lovelock:1971yv,Cai:2001dz}.
   The computation of the corresponding density of states would allow us to check whether the scaling of the micro-canonical entropy with the energy is affected by the addition of higher-curvature terms. Moreover, we expect that multiple central charges enter the generalization of the Cardy formula.
   We plan to compute the density of states associated to the R\'{e}nyi entropies in this setting in the future, including the case of shape deformations \cite{Hung:2011nu,Bianchi:2016xvf}. 
   \item \textbf{Relation to quantum complexity.}
   Another relevant quantity in the context of quantum information is computational complexity, which heuristically determines the optimal way to perform a certain unitary operation (\eg see the review \cite{Chapman:2021jbh}).
   Complexity has been related to measures of entanglement entropy in various ways, \eg \cite{PhysRevLett.127.020501,Carmi:2016wjl,Abt:2017pmf,Abt:2018ywl,Auzzi:2019vyh}. 
   In particular, the notion of binding complexity associates non-trivial cost to (non-local) gates acting among different parts of a system, and vanishing cost to (local) gates acting within a subregion \cite{Balasubramanian:2018hsu}. 
   In this context, one can show that binding complexity for a bipartite system can be expressed as a function of the Schmidt coefficients of a reduced density matrix \cite{Baiguera:2023bhm}.
   In the limit of a large number of Schmidt coefficients, the density of states plays an important role to properly define a complexity norm. It would be interesting to apply the results derived in the present paper to better understand the behaviour of binding complexity in the continuous limit, relevant for QFT applications. 
\end{enumerate}

%% file: Draft_Sections/App_Laplace.tex
\section{Mathematical tools}
\label{app:inverse_Laplace}

In this appendix, we collect some relevant properties of the inverse Laplace transform and of the saddle point approximation, which are the main analytic tools used to perform the method described in section~\ref{ssec:DOS_from_Renyi}.
First, we review in appendix~\ref{secInvLaplace} the analytic properties of the inverse Laplace transform in the complex plane.
Second, we discuss in appendix~\ref{ssec:saddle_point} the validity of the saddle point approximation.
Finally, we present in appendix~\ref{ssec:dominant_saddles} an explicit proof of the dominant saddle point for the case of holographic CFTs dual to an uncharged hyperbolic black hole.

\subsection{Inverse Laplace transform}
\label{secInvLaplace}

We recall that the Laplace transform of a continuous function $h(t)$ with $t \geq 0$ is defined by (\eg see section XIV--2 of reference \cite{gamelin2013complex})
\be
H(s) = \int_0^{\infty} dt \, e^{-t s} h(t) \, ,
\label{eq:Laplace_transform}
\ee
where $s \in \mathbb{C}$.
In order to make the integral absolutely convergent over the complex plane, we require that the function $h(t)$ is exponentially bounded as
\be
|h(t)| \leq M e^{B t} \, , \qquad
0 \leq t < \infty \, ,
\label{eq:convergence_smallf}
\ee
where $B,M$ are real constants.
In this way, the integral ~\eqref{eq:Laplace_transform} defines an analytic function $H(s)$ in the region of the complex plane delimited by $\Re(s) > B$.

To define the inverse Laplace transform, we rewrite eq.~\eqref{eq:Laplace_transform} as follows
\be
H(r-i p) = \int_{-\infty}^{\infty} dt \, e^{i p t} \left[  \Theta(t) h(t) e^{-t r} \right]  \, ,
\ee
where we performed the change of variables $s=r-i p$ (with $r, p \in \mathbb{R}$), and we inserted the Heaviside distribution $\Theta$ to extend the region of integration to the whole real axis. 
In this way, the expression formally takes the form of a Fourier transform, whose inverse is given by
\be
\Theta(t) h(t) e^{-t r} = \frac{1}{2 \pi} \int_{-\infty}^{\infty} dp \, e^{-i p t} H(r- i p) \, .
\ee
After inverting the previous change of variables, and performing simple algebraic manipulations, we find the inverse Laplace transform
\be
\Theta(t) h(t) = \frac{1}{2 \pi i} \int_{r-i \infty}^{r+i \infty} ds \, e^{s t} H(s) \, .
\label{eq:inverse_Laplace}
\ee
Notice that by construction, the inverse Laplace transform is only defined for $t \geq 0,$ as it was indeed required for the original function $h(t)$.
The integration is performed along a vertical line with $\Re(s) =r$, where $r$ is located to the right of all the singularities of $H$.

\begin{figure}[ht]
    \centering
    \includegraphics[scale=0.22]{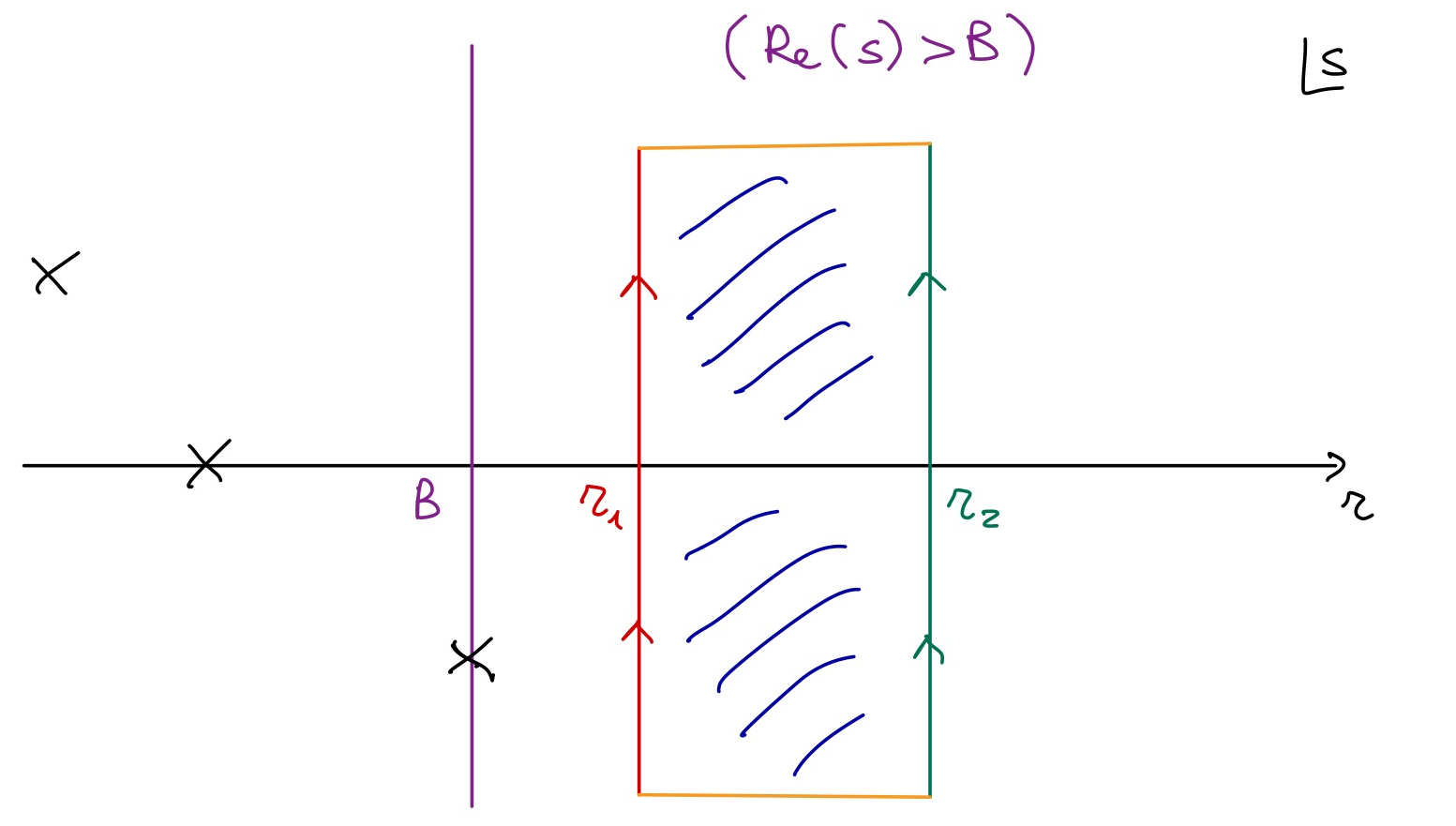}
    \caption{Schematic depiction of the inverse Laplace transform \eqref{eq:inverse_Laplace}. The inverse transform is only defined when $\Re(s) > B$, on the right of all the singularities of $H$. Due to Cauchy's theorem, the integral along any closed curve inside of which the function is analytic, vanishes. In the picture, crosses denote singularities of $H$. Therefore, we can deform the integration contour where the inverse Laplace transform is evaluated from the vertical red line to the green one (assuming that the integrals on the orange curves vanish).}
    \label{fig:Laplace}
\end{figure}

Given the properties of convergence of $h(t)$ required in eq.~\eqref{eq:convergence_smallf}, the inverse Laplace transform is an analytic function in the region  $\Re(s) > B$ only.
For this reason, as long as the integration contour is defined on the right of all the singularties of $H$, we can deform its location in the complex plane using Cauchy's theorem, as depicted in fig.~\ref{fig:Laplace}.

\subsection{Saddle point approximation}
\label{ssec:saddle_point}

We discuss the validity of a saddle-point approximation for the inverse Laplace transform of a function $H$.
The saddle point technique (also known as the method of steepest descent, see \cite{Fedoryuk1989} for a reference) consists in the approximation of an integral by deforming the integration contour to pass near a set of stationary points $s_*^{(j)}$, \ie defined by the necessary condition 
\be
\frac{\partial}{\partial s} \le  e^{st} H(s) \ri\Big|_{s=s_*^{(j)}} = 0 \, .
\ee
In order to apply the saddle point approximation, the following conditions should be met:
\begin{itemize}
    \item The saddle points lie in the range where $h(t)$ in eq.~\eqref{eq:inverse_Laplace} is analytic (\ie its Taylor expansion exists). In other words, $\Re(s_*^{(j)}) > B$, where $B$ is the real value of the rightmost singularity for the function $H$. 
    \item The integrand takes the form $H(s)= e^{\lambda g(s)}$ with $\lambda \rightarrow \infty$. 
    In the case of interest for this paper in eq.~\eqref{invLap}, such requirement is fulfilled because the function $H$ of interest is the holographic partition function $Z_n = e^{(1-n) S_n}$, where $S_n \propto 1/G_N$ and we consider the classical limit $G_N \rightarrow 0$.
    \item The saddle points are isolated and non-degenerate.
\end{itemize}

If these conditions are satisfied, then we can use Cauchy's theorem to deform the integration contour such that it passes through all the saddle points located in the region where the integrand is analytic.
Therefore, the method of steepest descent approximates the function $h(t)$ in eq.~\eqref{eq:inverse_Laplace} as follows \cite{Fedoryuk1989}
\be
h(t) \approx \sum_{j \, : \, \Re(s_*^{(j)})>B} \exp \le s_*^{(j)} t \ri \, \frac{e^{\lambda g(s_*^{(j)})}}{\sqrt{ 2\pi g''(s_*^{(j)})}} \, ,
\label{eq:app:sum_saddles}
\ee
where we used the parametrization $H(s)=e^{\lambda g(s)}$.
After identifying $e^{st} H(s) \rightarrow e^{f(n)}$ with the notation used in eq.~\eqref{invLap}, the previous formula reduces to the result \eqref{DOSf}.

The expression \eqref{eq:app:sum_saddles} contains a summation over all the saddle points in the region where $H$ is analytic.
Since this result is obtained in the limit $\lambda \rightarrow \infty$, it turns out that there is a preferred saddle point $\hat{s}_*$ in the main text which dominates the summation, thus allowing to approximate the result as
\be
h(t) \approx \frac{e^{\hat{s}_*t + \lambda g(\hat{s}_*)}}{\sqrt{2\pi g''(\hat{s}_*)}} \, .
\ee
In the main text we often make use of this latter approximation to evaluate the dominant contribution at high energies to the density of states.
We discuss the dominant saddle and subtleties regarding the remaining eligible saddles in appendix~\ref{ssec:dominant_saddles} below.

\subsection{The dominant saddle}
\label{ssec:dominant_saddles}

In the computation of the saddle points for the inverse Laplace transform of the R\'{e}nyi entropies for a holographic CFT dual to an uncharged black hole in section~\ref{ssec:computation_DOS}, we have found $d$ solutions \eqref{eq:nstar_highE_general}, from which only those satisfying \eqref{maxConstraint} are eligible. 
In principle, the density of states is obtained as the sum \eqref{eq:leading_DOS} over all the eligible saddle points, as we motivated in eq.~\eqref{eq:app:sum_saddles}. 
Here, we determine which contributions in the summation are dominant.

In the high energy limit, all but one saddle point arise in pairs, \ie two solutions share the same real part.
There is a unique real saddle, \ie a saddle without a partner, whose density of states is discussed in \eqref{eq:leading_DOS_largeE}. The other saddles require more care; we refer to them as complex saddles in this appendix. Given the fact that all saddles \eqref{eq:nstar_highE_general} approach each other in the regime \eqref{eq:assumptions_cutoff}, we have to first check that the saddle point approximation does not break down, as would happen, if the Bell curves surrounding two saddles overlap.

In order to achieve this, we check that the distance between two saddles with identical real part, \ie
\begin{equation}
 \left| n_*^{(k)}-n_*^{(-k)} \right|
 =
 \frac{4}{d}\sin\le 2\pi\frac{k}{d}\ri \le \frac{L^{d-2}}{m(E)} \ri^{\frac{1}{d}}
 \approx
  \frac{4}{d}\sin\le 2\pi\frac{k}{d}\ri \le \frac{d-1}{\cE(d)}\frac{E_0}{E-E_0} \ri^{\frac{1}{d}} \, ,
\end{equation}
is larger than the standard deviation
\begin{align}
    \sigma
    =
    \le 2\pi |f''(n_*^{(k)})| \ri^{-1/2}
    =
    \le \pi d^2\le \frac{\cE(d)(E-E_0)^{d+1}}{(d-1)E_0}\ri^{\frac{1}{d}} \ri^{-1/2} 
\end{align}
of the Gaussian integrand in the first line of \eqref{DOSf}. 
Their ratio 
\begin{align}
    \frac{\sigma}{\left|n_*^{(k)}-n_*^{(-k)} \right|}
    \propto\frac{1}{\sqrt{E_0}}\le \frac{E-E_0}{E_0} \ri^{-\frac{1}{2} \le 1 - \frac{1}{d} \ri }
    \ll1
\end{align}
implies that the distance is larger than the width of the normal distributions in the integrand, hence corroborating that \eqref{eq:leading_DOS} is still a good approximation.

Complex saddles can satisfy eq.~\eqref{maxConstraint} in dimensions $d\geq5$.
Since they always appear in complex conjugate pairs, their density of states becomes 
\begin{align}
    n_*^{(k)}\&\,n_*^{(-k)}:
    \quad
    D_{(k)}(E)
    =
    2\Re\biggl\{
    &\exp[2\le \frac{E_0}{\cE(d)}\ri^{\frac{1}{d}}\left(\frac{E-E_0}{d-1}\right)^{\frac{d-1}{d}}e^{-2\pi ik/d}]\notag\\
    &\times\biggl(
    \pi d^2\le \frac{\cE(d)(E-E_0)^{d+1}}{(d-1)E_0}\ri^{\frac{1}{d}}\,e^{2\pi i k/d}
    \biggr)^{-\frac{1}{2}} 
    \biggr\}\, .
    \label{eq:DOSotherSaddles}
\end{align}
Given that complex conjugates are simply summed in this expression, it suffices to restrict the phase in \eqref{eq:nstar_highE_general} to the first quadrant $0<2\pi k/d<\pi/2$, in line with \eqref{maxConstraint}.

It is now interesting to investigate whether the density of states \eqref{eq:DOSotherSaddles} or \eqref{eq:leading_DOS_largeE} is dominant. To that end, it is convenient to introduce the shorthand notation
\begin{equation}
    \Lambda(E)
    =
    2\le \frac{E_0}{\cE(d)}\ri^{\frac{1}{d}}\left(\frac{E-E_0}{d-1}\right)^{\frac{d-1}{d}}\,,
    \qquad
    \lambda(E)
    =
    \pi d^2\le \frac{\cE(d)(E-E_0)^{d+1}}{(d-1)E_0}\ri^{\frac{1}{d}} \, ,
\end{equation}
which are always positive. Temporarily dressing the density of states \eqref{eq:leading_DOS_largeE} with a subscript $(0)$, our candidate densities can thus be written compactly as
\begin{equation}
    D_{(0)}(E)
    =
    \frac{1}{\sqrt{\lambda(E)}}e^{\Lambda(E)}\,,
    \qquad
    D_{(k)}(E)
    =
    2\Re\biggl\{ \frac{1}{\sqrt{\lambda(E)e^{2\pi i k/d}}}e^{\Lambda(E)e^{-2\pi i k/d}} \biggr\} \, .
\end{equation}
The second of these two expressions is easily manipulated into the form
\begin{equation}
    D_{(k)}(E)
    =
    \frac{2}{\sqrt{\lambda}}
    \exp \biggl[
        \Lambda\cos\le 2\pi \frac{k}{d} \ri
        \biggr]
    \cos \biggl[    
        \Lambda\sin \le 2\pi \frac{k}{d} \ri + \pi \frac{k}{d}
        \biggr] \, .
\end{equation}
Both densities can now be compared
\begin{equation}
 \frac{D_{(k)}(E)}{D_{(0)}(E)}
 =
 2\exp \biggl[
        -\Lambda\le 1-\cos\le 2\pi \frac{k}{d} \ri \ri
        \biggr]
    \cos \biggl[    
        \Lambda\sin \le 2\pi \frac{k}{d} \ri + \pi \frac{k}{d}
        \biggr] \, .
\end{equation}
The cosine is bounded and the argument of the exponential is always negative. Given that $\Lambda$ is very large, the ratio is suppressed at large energies. The dominant saddle is hence always the partner-less real saddle $n_*^{(0)}$. 

We note that very small $k/d$ can counter the effect of large $\Lambda$ to some degree in the exponential.  In this work we contend ourselves with always choosing a large enough energy $E$ (and thus $\Lambda$) such that these issues are avoided. This is possible because for fixed $d$ the angle $k/d$ is bounded from below.

The obvious question whether phase transitions are encountered at lower energies arises. We leave the exploration of this interesting issue to future work, see also section \ref{sec:Future} for more comments.

%% file: Draft_Sections/App_Universality.tex
\section{Universality of the functional dependence on the energy}
\label{app:universality_Edep}

In order to clarify which part of the density of states is independent of the regularization scheme, let us re-express the objects we computed in the main text.
Notice that the dependence of the R\'{e}nyi entropies on the cut-off scale factorizes, and can be expressed as
\be
S_n = E_0 \tilde S_n 
\ee
where $E_0$ contains all of the cutoff dependence of $S_n$.
This fact can be seen from the holographic result \eqref{eq:charged_Renyi_holo}, together with the definition of $E_0$ in the uncharged \eqref{eq:def_E0} and SUSY cases \eqref{eq:def_E0_SUSY}.
Correspondingly, the partition function reads
\be
Z_n = e^{(1-n) S_n} = e^{E_0(1-n) \tilde S_n}.
\ee
So far, the previous objects are the same as in the main text. Let us now re-express $E=E_0 \tilde E$, and compute the density of states $D(E_0 \tilde E)$ with the inverse Laplace transform,
\be
D(E_0 \tilde E) = \int_\mathcal{C} dn e^{E_0(1-n) \tilde S_n + n E_0 \tilde E} = \int_\mathcal{C} dn e^{E_0 \tilde f(n)}
\ee
where the contour is defined below eq.~\eqref{invLap}.

The saddle point $n_*(\tilde E)$ is given by
\be
E_0 \tilde f'(n)=0
\ee
which is independent of $E_0$.
The method of steepest descent now gives
\be
D(E_0 \tilde E) = \frac{1}{\sqrt{2 \pi E_0 \tilde f''\left(n_*(\tilde E)\right)}}e^{E_0 \tilde f\left(n_*(\tilde E)\right)},
\ee
where $\tilde f(n_*(\tilde E))$ and $\tilde f''(n_*(\tilde E))$ are cut-off independent.  

Notice that if the modular Hamiltonian of the reduced density matrix is rescaled as $\rho_A = e^{-H} = e^{-E_0 \tilde H}$,
then we get
\be
Z_n = \int_0^\infty d\tilde E \, g(\tilde E)e^{-n E_0 \tilde E} = \int_0^\infty dE \, \frac{ g(E/E_0)}{E_0}e^{-n E} = \int_0^\infty dE D(E)e^{-n E},    
\ee
which defines the relation between $D$ and $g$, the densities of states of $H$ and $\tilde H$, respectively. Specifically, 
\be
D(E_0 \tilde E) = \frac{g(\tilde E)}{E_0}.
\ee
From this relation we obtain the micro-canonical entropy of $\tilde H$:
\be
S(\tilde E) = \log(g(\tilde E)) = E_0 \tilde f \left(n_*(\tilde E)\right) + \frac{1}{2}\left[\log(E_0) - \log(2 \pi \tilde f''\left(n_*(\tilde E)\right))\right]
\ee
This formula shows that the leading contribution $\tilde{f}(n_*)$ has a universal functional dependence on the energy $\tilde{E}$, while all the scheme dependence is contained in the prefactor $E_0$.
Similarly, the functional dependence on $E$ of the subleading logarithmic contribution in $f''(n_*)$ is universal.